\documentclass[12pt]{article}
\usepackage{epsfig,graphicx,amssymb,bm,amsmath}

\def \Zbar {\overline{Z}}
\def \be {\begin{equation}}
\def \ee {\end{equation}}
\def \bea {\begin{eqnarray}}
\def \eea {\end{eqnarray}}

\def \R {{\textsf{I}\kern-.10em \textsf{R}}}
\def \T {{\textsf{T}\kern-.45em \textsf{T}}}
\def \C {{\textsf{C}\kern-.37em \textsf{C}}}
\def \Z {{\textsf{Z}\kern-.35em \textsf{Z}}}
\def \H {{\textsf{I}\kern-.10em \textsf{H}}}
\def \S {{\textsf{S}\kern-.37em \textsf{S}}}

\def \dels {\partial\kern-.5em / \kern.5em}
\def \As {{A\kern-.5em / \kern.5em}}
\def \Ds {D\kern-.7em / \kern.5em}

\newcommand{\ena}{\end{eqnarray}}

\def\bbox{{\,\lower0.9pt\vbox{\hrule \hbox{\vrule height 0.2 cm
\hskip 0.2 cm \vrule height 0.2 cm}\hrule}\,}}
\newcommand{\dsl}{\pa \kern-0.5em /}

\newcommand{\pa}{\partial}

\def \K {{\tt I\kern-.25em K}}

\begin{document}


\begin{titlepage}
\begin{center}

\hfill\parbox{4cm}{
{\normalsize\tt hep-th/0506023}}\\

\vskip .5in

{\LARGE \bf Weyl Card Diagrams}

\vskip 0.3in

{\large Gregory C. Jones$^a$\footnote{\tt
  jones@physics.harvard.edu} and
John E.\ Wang$^{b}$\footnote{{\tt hllywd2@phys.ntu.edu}} }

\vskip 0.15in

${}^a$ {\it Department of Physics, Harvard University, Cambridge, MA
02138}\\[3pt]
${}^b$ {\it Department of Physics, National Taiwan University \\
Taipei 106, Taiwan}\\
[0.3in]

\end{center}

\vskip .3in

\begin{abstract}
\normalsize\noindent To capture important physical properties of a
spacetime we construct a new diagram, the card diagram, which
accurately draws generalized Weyl spacetimes in arbitrary
dimensions by encoding their global spacetime structure,
singularities, horizons, and some aspects of causal structure
including null infinity.  Card diagrams draw only non-trivial
directions providing a clearer picture of the geometric features
of spacetimes as compared to Penrose diagrams, and can change
continuously as a function of the geometric parameters. One of our
main results is to describe how Weyl rods are traversable horizons
and the entirety of the spacetime can be mapped out.  We review
Weyl techniques and as examples we systematically discuss
properties of a variety of solutions including Kerr-Newman black
holes, black rings, expanding bubbles, and recent spacelike-brane
solutions. Families of solutions will share qualitatively similar
cards. In addition we show how card diagrams not only capture
information about a geometry but also its analytic continuations
by providing a geometric picture of analytic continuation.  Weyl
techniques are generalized to higher dimensional charged solutions
and applied to generate perturbations of bubble and S-brane
solutions by Israel-Khan rods. This paper is a condensed and
simplified presentation of the card diagrams in hep-th/0409070.

\end{abstract}

\vfill

\end{titlepage}
\setcounter{footnote}{0}

\pagebreak
\renewcommand{\thepage}{\arabic{page}}
{\baselineskip=5mm\tableofcontents}


\section{A new diagram for spacetime structure}

Spacetimes are geometrical objects, independent of the coordinates
with which we describe them.  However, spacetimes are typically
presented and visualized in a specific coordinate system. If the
coordinates are poorly chosen, many properties of the spacetime
such as horizons, causally connected spacetime points, maximal
extensions and null infinity are not readily apparent.

A simplification occurs if a $D$ dimensional Lorentzian spacetime
has enough fibered directions (like a $(D-2)$-sphere) or other
ignorable directions.  One can draw two dimensional diagrams for
the remaining directions and such Lorentzian $-+$ signature
spacetime slices can be conformally compactified leading to
Penrose diagrams.

Penrose diagrams are quite useful in understanding spacetime
geometry and successful especially in understanding causal
structure although there are some limitations to this approach.
For instance just knowing the Penrose diagram for the subextremal
$Q^2<M^2$ Reissner-Nordstr\o m black hole does not tell us what
happens to the spacetime structure in the chargeless or extremal
limits.   For more complicated spacetimes, Penrose diagrams (which
assume symmetry or fibering) can only draw a slice of the
spacetime.  As a known example, the Penrose diagram for a Kerr
black hole does not clearly depict the ring singularity and the
possibility of crossing through the interior of the ring into a
second universe. In addition, recently analytic continuation has
been applied to black hole solutions to yield bubble-type
\cite{Witten:1981gj,Aharony:2002cx,
Birmingham:2002st,Balasubramanian:2002am,Biswas:2004xc,Astefanesei:2005eq}
or S-brane \cite{Astefanesei:2005eq,
Gutperle:2002ai,roll,ChenYQ,KruczenskiAP,sugraSbranes,Strotalk,Ohta:2003uw,
Maloney:2003ck,Jones:2004rg,Wang:2004by,Tasinato:2004dy,Lu:2004ye,
Gutperle:2004vh,Jones:2004pz} solutions.  Oftentimes this is done
in Boyer-Lindquist type coordinates which are hard to visualize.
Again we are not left with a clear picture of the resulting
spacetime and the Penrose diagrams are missing important
noncompact spatial directions.

It is useful to have an alternative diagram which can also capture
important features of a spacetime. For this reason in this paper
we expand the notion of drawing spacetimes in Weyl space
\cite{Myers:rx,EmparanWK}.  Because our diagrams have the
appearance of playing cards glued together we will dub them Weyl
card diagrams.

To understand the construction of a card diagram we recall that in
$D=4$ dimensions a Weyl solution in canonical coordinates
\cite{EmparanWK,Weylpaper,EmparanBB} is written as $$ ds^2=-f
dt^2+f^{-1}[e^{2\gamma}(d\rho^2+dz^2)+\rho^2 d\phi^2] $$ where $f$
and $\gamma$ are functions of $\rho, z$. The original Weyl class
requires two commuting orthogonal Killing fields $\partial_t,
\partial_\phi$ in four dimensions \cite{Weylpaper}, or $D-2$
fields for general $D$ dimensions \cite{EmparanWK}.\footnote{
Non-Weyl, axisymmetric spacetimes in $D\geq 4$ are discussed in
\cite{Myers:rx,Gregory}.} Sometimes Weyl solutions are called
axially-symmetric gravitational solutions although they in fact
are more general. We also include the Weyl-Papapetrou class for 2
commuting Killing vectors in $D=4$ \cite{papapetrou}, and allow
charged static solutions in $D\geq 4$ (see the Appendix to this
paper). Furthermore stationary vacuum solutions in $D\geq 4$ are
covered with the recent work of \cite{Harmark:2004rm}. In four and
five dimensions this generalized Weyl class includes spinning
charged black holes and rings
\cite{EmparanWK,Harmark:2004rm,Emparan:2001wn,Elvang:2003yy,
Elvang:2003mj,Emparan:2004wy,Elvang:2004rt,Elvang:2004ds,Elvang:2004xi}
as well as various arrays
\cite{Jones:2004rg,Jones:2004pz,IsraelKhan} of black holes,
spacelike-branes, and includes backgrounds like Melvin fluxbranes
\cite{Jones:2004rg,Jones:2004pz,
Melvin,Dowker:1995gb,zeroduality,Emparan:2001gm,Cornalba:2002fi}
and spinning ergotubes \cite{Siklos}.

When constructing card diagrams, we will draw only Weyl's
canonical coordinates $(\rho,z)$, or coordinates related to them
via a conformal transformation.  The Killing coordinates are
ignorable and so this diagram is efficient and will show all
details of the spacetime. Since there are only two nontrivial
coordinates, card diagrams are two dimensional and easy to draw
like Penrose diagrams.  The difference however is that while
Penrose diagrams are truly two-dimensional, card diagrams are
drawn as if embedded in three dimensions.  When a $(\rho,z)$
region of the spacetime has Euclidean $++$ signature, we draw the
two coordinates $(\rho,z)$ horizontally; and this makes a
horizontal card.  For Lorentzian signature $-+$ regions we use
$(\rho',z)$ or $(\tau,\rho)$, and draw the timelike coordinate
vertically; this makes a vertical card. Causal structure is
automatically built into the vertical cards since for example the
directions $(\tau,\rho)$ appear conformally only through the
combination $-d\tau^2+d\rho^2$. Horizontal cards and vertical
cards are attached together at Killing horizons and so card
diagrams resemble a gluing-together of a house of playing cards.

In this paper we will present card diagrams for the familiar
spacetimes of black holes, as well as expanding bubbles, S-branes,
and black rings.  Many other spacetimes including the S-dihole,
infinite periodic universe, C-metric, and multiple-rod solutions
in 4 and 5 dimensions are presented in \cite{Jones:2004pz}, and
spacetimes derived from 4 and 5 dimensional Kerr geometries will
be presented in \cite{joneswangfuture}.

In Section 2 we review the Schwarzschild black hole in the usual
coordinates and in Weyl coordinates.  By extending through the
horizon and properly representing the interior of the black hole
we construct the first card diagram. We emphasize the construction
of the interior of the black hole as a vertical card comprised of
four triangles unfolded across special null lines.  We then
discuss general card diagram properties such as null lines, list
the available card types, and geometrically describe the
$\gamma$-flip analytic continuation procedure.

In Section 3, we discuss the sub/super/extremal Reissner-Nordstr\o
m black hole card diagrams, the Kerr black hole, and the black
ring/C-metric card diagrams.  We then show that a spacetime can
have multiple card diagrams and as examples present the elliptic,
hyperbolic, and parabolic representations of the charged Witten
bubble and charged Spacelike brane which we also call
S-Reissner-Nordstr\o m. Finally, as a newer example we discuss the
(twisted) S-Kerr solution \cite{Wang:2004by,Tasinato:2004dy}.

We conclude with a discussion in Section 4.  We give an appendix
on perturbing Witten bubbles and S-branes by introducing
Israel-Khan rods, in their hyperbolic or elliptic representations.
We also give an appendix on how the higher dimensional vacuum Weyl
Ansatz can be extended to include electromagnetic fields.

This paper is a condensed presentation of the card diagrams in
\cite{Jones:2004pz}.

\section{Schwarzschild and general card diagrams}

In this section we review the Schwarzschild black hole as an
example of a Weyl spacetime and then we explain the construction
of its associated Weyl card diagram.  General features and
properties of card diagrams are also developed.

Up to now if a solution such as the Schwarzschild black hole had
horizons, then only the regions outside the horizons have been
drawn in Weyl coordinates \cite{EmparanWK,EmparanBB}. To go
through a nonextremal horizon, the Weyl coordinate $\rho$ must be
allowed to take imaginary values.  We discuss how the horizon can
be represented as a junction of four regions which we call four
cards.  The regions outside the horizon will be drawn as two
horizontal cards while the regions between the horizon and the
singularity will be drawn as two vertical cards.  The interior
vertical cards naively are problematic and have fourfold-covered
triangles bounded by `special null lines'.  However the triangles
can be unfolded and glued together into a square along the
`special null lines' to achieve a singly covered representation of
the spacetime in Weyl coordinates by properly choosing branches of
a square root in the solution.

Note that card diagrams represent the spaces on which we solve the
Laplace equation (horizontal card) or wave equation (vertical
card) to find a Weyl metric.  For example, the Schwarzschild black
hole has a finite length uniform density rod source along the
$z$-axis generating the potential $\log f$, and the remainder of
its $z$-axis encodes the vanishing of the $\phi$-circle.  Thus
card diagrams give a full account of the boundary conditions
necessary to specify the spacetime.

\subsection{Schwarzschild black holes}

\begin{figure}[htb]
\begin{center}
\epsfxsize=6.5in\leavevmode\epsfbox{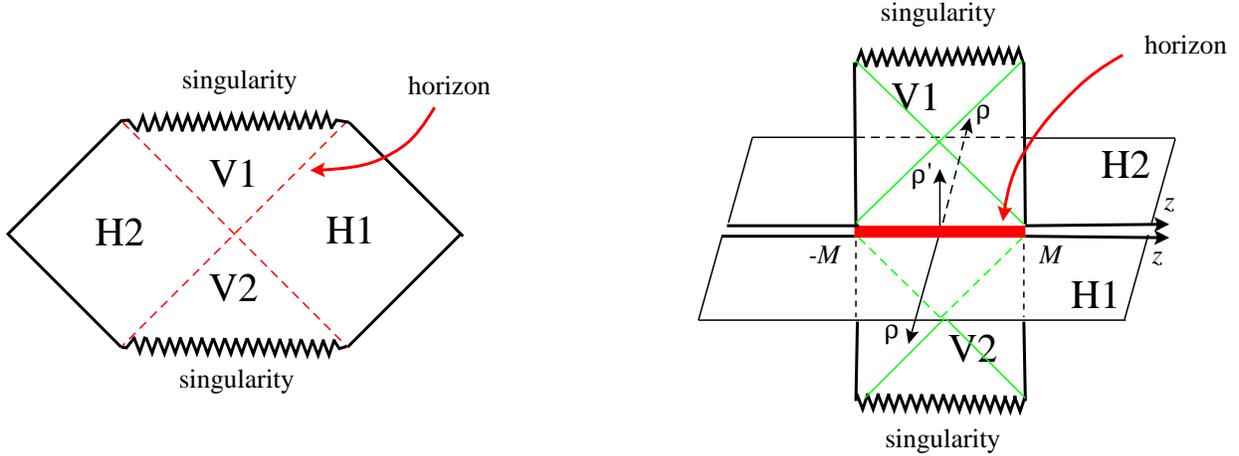}
\caption{The Schwarzschild black hole represented both as a
Penrose diagram as on the left and as a Weyl card diagram on the
right.  The cards V1 and V2 are vertical squares made of four
triangles, while the cards H1 and H2 are horizontal and are
infinite half-planes. The four cards join together along the black
hole horizon on the $z$-axis.} \label{Penrose-Weyl-comparison}
\end{center}
\end{figure}

This section will describe the construction of the Schwarzschild
black hole card diagram.  The Penrose and card diagrams are
compared in Fig.~\ref{Penrose-Weyl-comparison}.

\begin{figure}[htb]
\begin{center}
\epsfxsize=7cm\leavevmode\epsfbox{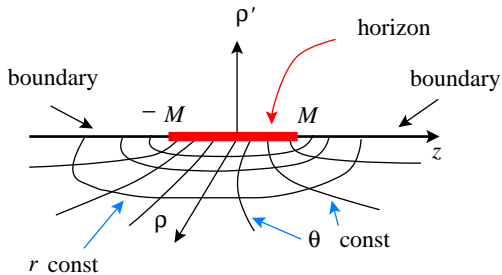} \caption{In Weyl
coordinates the Schwarzschild black hole is typically represented
as a $\rho, z$ horizontal half-plane showing one external
universe, H1, of $r\geq 2M$.   The horizon is represented as a rod
along $\rho=0$, $-M\leq z\leq M$.  The $0<r<2M$ region will be
drawn along the vertical perpendicular $\rho^\prime$-direction.}
\label{weylrod}
\end{center}
\end{figure}

The Schwarzschild metric in spherically symmetric (Schwarzschild)
coordinates is
\begin{equation}\label{schwarz}
ds^2=-(1-2M/r)dt^2+(1-2M/r)^{-1}dr^2+r^2d\theta^2+r^2\sin^2\theta
d\phi^2 \ .
\end{equation}
There is a horizon at $r=2M$ and a curvature singularity at $r=0$.
On the other hand, in Weyl's canonical coordinates the metric
ansatz is \cite{EmparanWK,Weylpaper,EmparanBB} it is
\begin{equation}ds^2=-f
dt^2+f^{-1}(e^{2\gamma}(d\rho^2+dz^2)+\rho^2
d\phi^2)\end{equation} where $f$ and $\gamma$ are functions of the
coordinates $\rho$ and $z$:
\begin{eqnarray}
f&=&{(R_++R_-)^2-4M^2\over (R_++R_-+2M)^2},\label{schwarzschildweylrep}\\
e^{2\gamma}&=&{(R_++R_-)^2-4M^2\over 4R_+R_-},\nonumber\\
R_{\pm}&=&\sqrt{\rho^2+(z\pm M)^2}.\nonumber
\end{eqnarray}
Previously attention was restricted to the H1 half-plane $\rho\geq
0$, $-\infty<z<\infty$, known as Weyl space, which describes the
exterior of the black hole and whose horizon is represented by a
`rod' line segment $\rho=0$, $-M\leq z\leq M$; see
Fig.~\ref{weylrod}. Note that the non-Killing 2-metric is
conformal to the Euclidean flat space $d\rho^2+dz^2$. The
coordinate transformation between Schwarzschild and Weyl
coordinates is
\begin{eqnarray}\label{coordinatetrans}
\rho&=&\sqrt{r^2-2Mr}\,\sin\theta,\\
z&=&(r-M)\cos\theta.\nonumber
\end{eqnarray}

Now we wish to ask how Weyl's coordinates draw the spacetime
inside the horizon.  The Schwarzschild coordinates
(\ref{coordinatetrans}) tell us that for $0<r<2M$, $\rho$ is
imaginary and so we set $\rho'=i\rho$. (In general we must perform
an analytic continuation of Weyl coordinates to go through a
horizon which are at the zeros of the Weyl functions $f,
e^{2\gamma}$.)  The analytic continuation gives a region with a
conformally Minkowskian metric $-d\rho'^2+dz^2$ and we will draw
this region as being vertical and perpendicularly attached to the
horizontal card at the horizon $-M\leq z\leq M$. The vertical
direction is always timelike in card diagrams.

Of course the Schwarzschild horizon structure is more complicated
than just having two cards joined together.  For example we know
the Penrose diagram in the $t,r$ coordinates is divided into four
regions that meet in a $\times$-horizon structure.  In Weyl
coordinates, we already saw that we can extend $\rho\rightarrow
\pm i \rho'$ or go to negative $\rho$. This gives us the four
regions of the Schwarzschild black hole in Weyl coordinates.  Two
regions will be horizontal at real values of $\rho$ and two
regions will be vertical with imaginary values of $\rho$.  So in
addition to the first horizontal card in front of the horizon, we
also have a copy of the horizontal external universe behind the
horizon and attach two vertical cards, one above and one mirror
copy below the horizontal cards (see
Fig.~\ref{Penrose-Weyl-comparison}).  All together, four different
regions attach together at the same $-M\leq z\leq M$ rod horizon.

The four regions labelled H1, H2, V1 and V2 in the Penrose diagram
map to the similarly labelled four regions on the card diagram in
Fig.~\ref{Penrose-Weyl-comparison}. Note that the Weyl cards draw
the $r,\theta$ coordinates which is different from the $t, r$
coordinates of the Penrose diagram.  However the fact that the
radial coordinate $r$ describes four distinct regions, two where
$\partial_r$ is spacelike and two where it is timelike, is still
apparent in the Weyl card diagram.  So while a Penrose diagram
always has a Lorentzian $-+$ signature, a card diagram will flip
from being Euclidean $++$ to Lorentzian $-+$ across a nonextremal
Killing horizon.

\begin{figure}[tp]
\begin{center}
\includegraphics[width=8cm]{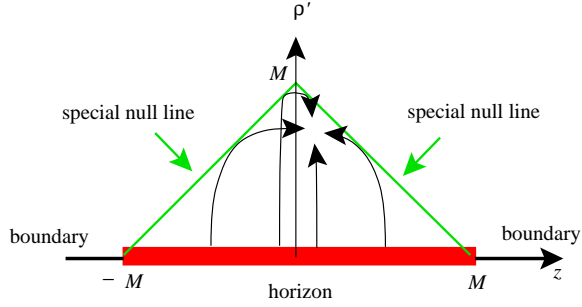}
\caption{The Weyl representation of the interior of the
Schwarzschild black hole naively gives a triangle with base length
$2M$ and height $M$.  As we illustrate, the triangle interior is
covered four times by orbits of $r$ at four different values of
$\theta$.} \label{4xcoverfig}
\end{center}
\end{figure}

Let us now examine the construction of the upper vertical card
extended in the $\rho^\prime,z$ directions. Looking at an
$r$-orbit for $r\in [0,2M]$  on the vertical card, we see that in
Weyl coordinates $0\leq \rho'\leq M\pm z$. The bounding lines
where $R_\pm=\sqrt{-\rho'^2+(z\pm M)^2}=0$, we will call `special
null lines,' and they are a general feature of vertical cards with
focal points (the rod endpoints $z=\pm M$). Here, special null
lines are the envelope of the $r$-orbits as we vary $\theta$.
Inside the horizon the Schwarzschild coordinates apparently fill
out a 45-45-90 degree Weyl triangle with hypotenuse length $2M$ a
total of four times, as shown in Figure~\ref{4xcoverfig}.

\begin{figure}[tp]
\begin{center}
\includegraphics[width=8cm]{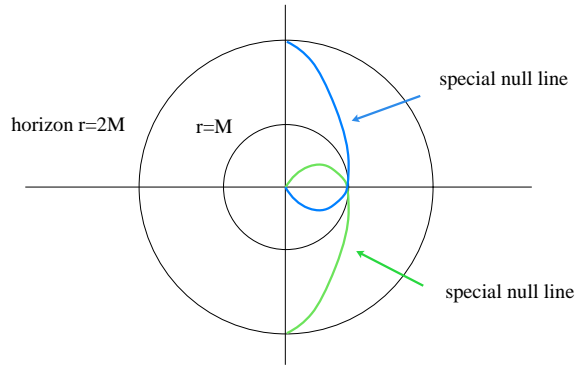}
\caption{The Weyl null lines for the Schwarzschild black hole
correspond to two 3-surfaces which partition $0<r<2M$ into
four subregions.}
\label{cardioids}
\end{center}
\end{figure}

Special null lines play an important role in Weyl card diagrams so
let us explain their significance. Keep in mind that we have
already broken the manifest spherical symmetry when we have
written the Schwarzschild metric in Weyl coordinates, so the
existence of preferred special null lines is relative to this
chosen axis. Consider the two 3-surfaces \be\label{BLR}
R_{\pm}=r-M\pm  M \cos\theta=0,\ee which are drawn in
Fig.~\ref{cardioids}. These surfaces bound the trajectories of
light rays that do not move in the Killing directions.  The
surfaces intersect at $r=M$ and partition the black hole interior
into four subregions. These regions correspond to the four Weyl
triangles.

It is clear from (\ref{BLR}) that $R_{\pm}$ is positive outside
the horizon and there is no difficulty going to negative values
inside the horizon. On the other hand in terms of Weyl
coordinates, the functions $R_\pm=\sqrt{-\rho'^2+(z\pm M)^2}$ are
the square root of a positive number when $\rho'<z\pm M$ and
imaginary if $\rho'>z\pm M$.  Clearly, instead of dealing with
imaginary values of $R_\pm$, the way to go `beyond' the special
null line $R_\pm=0$ is to keep $\rho'<z\pm M$ but use the other
square root branch for $R_\pm$ as it enters in the Weyl functions
$f$ and $\gamma$ by explicitly replacing $R_\pm \to -R_\pm$. Since
we can pass $R_+=0$ and/or pass $R_-=0$, it is clear that the four
different branches of the square root functions differentiate the
four copies of the Weyl triangle.

\begin{figure}[htb]
\begin{center}
\epsfxsize=3in\leavevmode\epsfbox{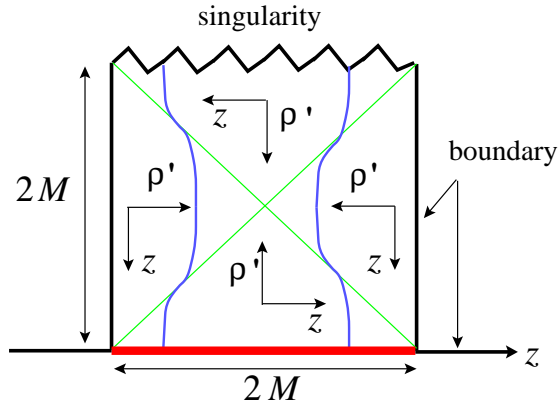}
\caption{Unfolding the four triangles along the $45^\circ$ special
null lines for $0<r<2M$, produces a square, V1, of side length
$2M$. The horizon is at the bottom and the singularity at the top.
We also draw two lines of constant Schwarzschild coordinate
$\theta$ and varying $r$.} \label{buildcard6}
\end{center}
\end{figure}

From (\ref{schwarzschildweylrep}), passing each $R_\pm=0$ null
line changes the sign of $e^{2\gamma}$ and hence exchanges the
timelike and spacelike nature of $\rho'$ and $z$.  Since the
vertical direction on a vertical card always represents time, two
of the triangles are drawn turned on their sides, their
hypotenuses vertical and so timelike. The four triangles
describing the interior of the black hole $0\leq r\leq 2M$ glue
together along the special null lines to fit neatly into a square;
see Fig. \ref{buildcard6}. Because of the unfolding of the
triangles, the positive $z$-direction on the top triangle (and any
attached horizontal cards) points in the opposite direction
compared to that on the original horizontal card.

The upper vertical cards is thus a square of length $2M$. The
bottom of the upper card V1 is the black hole horizon which
connects to three other cards in a four card junction. The right
and left edges of this vertical card correspond to $\theta=0,\pi$
and are the boundaries where $\rho^\prime=0$ the $\phi$-circle
vanishes. The top edge of the card represents the $r=0$ curvature
singularity. The second vertical card V2 is built in analogous
fashion except the square is built in a downwards fashion towards
negative values of $\rho'$.  Additionally there is a second
horizontal card plane, H2, identical to H1 at negative values of
the $\rho$ coordinate attached to the same horizon along $[-M,M]$
on the $z$-axis.

\begin{figure}[htb]
\begin{center}
\epsfxsize=4in\leavevmode\epsfbox{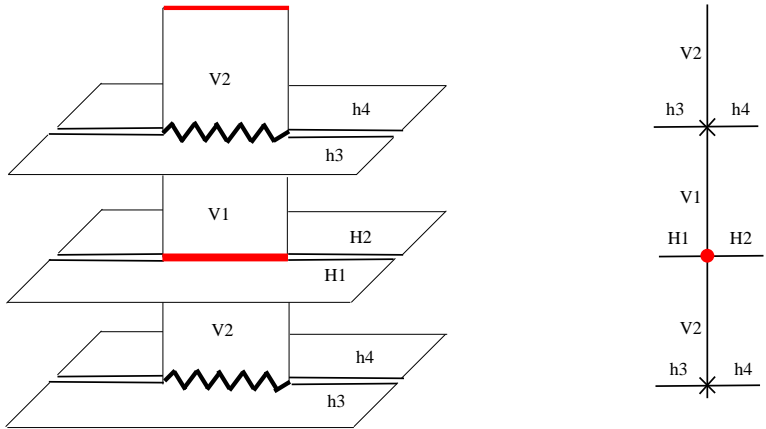}
\caption{The extended card diagram for Schwarzschild includes both
positive (H1,H2) and negative (h3,h4) mass universes half-planes.
Its cross-section for $z=0$ is drawn on the right.}
\label{FullSchwarzCardfig}
\end{center}
\end{figure}

One typically stops the construction of the Schwarzschild
spacetime with the above four regions, and considers the $r<0$
part of the metric to be a separate spacetime.  However for
reasons which become clear when we look at the Reissner-Nordstr\o
m and Kerr black holes in Sections~\ref{RNsubsub} and
\ref{Kerrsubsec}, we continue the card diagram past the
singularity and attach two horizontal half-plane corresponding to
negative-mass (or $r<0$) universes h3 \& h4, and further vertical
card above the singularity which is identical to V2. The two new
horizontal cards each represent negative mass-universes with no
horizon and a naked singularity along $-M\leq z\leq M$. Note that
although Penrose diagrams for the $r>0$ and $r<0$ regions of
Schwarzschild cannot attach together since the singularity is
spacelike in one region and timelike in another, cards can
naturally attach at this singularity.  The extended card diagram
for Schwarzschild, shown in Fig.~\ref{FullSchwarzCardfig}, is an
infinite array of repeated cards representing positive (H1,H2) and
negative (h3,h4) mass universes and inside-horizon regions.

\subsection{General properties of card diagrams}

Having described the construction of the card diagram for
Schwarzschild we now turn to general remarks about these card
diagrams.  Card diagrams can be constructed in general
D-dimensions starting from a generalized Weyl ansatz including the
charged cases discussed in Appendix \ref{EWeylapp}. Horizontal
cards are conformally Euclidean and represent stationary regions.
Vertical cards are always conformally Minkowskian and represent
regions with ($D-2$) spacelike Killing fields.  On a vertical card
time is always in the vertical direction and a spacetime point's
causal future lies between $45^\circ$ null lines on the card.
Weyl's coordinates certainly go bad at horizons, so these diagrams
are not a full replacement for Penrose diagrams at understanding
causal structure or particle trajectories. However, it is clear
for example from Fig.~\ref{Penrose-Weyl-comparison} that two
vertical and two horizontal cards attach together in a orthogonal
$+$-configuration in precisely the sense of the $\times$-horizon
structure of the Penrose diagram.

The prototypical horizon is
that of the two Rindler and two Milne wedges of flat space; this
spacetime has two horizontal half-plane cards and two vertical
half plane cards that meet along the horizon being the whole $z$-axis.
Zooming in on a non-extremal horizon of any card diagram yields
this Rindler/Milne picture.

A rod endpoint $(\rho,z)=(0,z_i)$ such as $z=\pm M$ for
Schwarzschild is a `focus' for the Weyl diagram and often
represents the edge of a black hole horizon or the end of an
acceleration horizon on a horizontal card.  Generally, multi-black
hole Weyl spacetimes depend only on distances to foci, such as
$R_\pm$ in the case of Schwarzschild \cite{sibgatullin}.

To understand that in general it is natural to change the branch
of the distance functions, $R_i$, when crossing the special null
line that emanates from the foci at $z=z_i$, take some Weyl
spacetime and imagine moving upwards in a vertical card to meet
the special null line where $R_i=0$ by increasing time $\rho'\geq
0$ for fixed spatial $z$. Rearranging
$R_i=\sqrt{-\rho'^2+(z+z_i)^2}$ as the semi-ellipse
$$R_i^2+\rho'^2=(z+z_i)^2,$$
we see that a smooth traversal of this semi-ellipse across $R_i=0$
requires a change in the sign of $R_i$.

In many of our solutions, special null lines are used to reflect
vertical triangular cards to create full, rectangular cards.
However, in the Bonnor-transformed S-dihole geometry of
\cite{Jones:2004pz} as well as double Killing rotated extremal
geometries and parabolic representations of the bubble and
S-Schwarzschild in Sec. \ref{thirdcard-Poincare}, the special null
lines will serve as conformal boundaries at null infinity ${\cal
I}^\pm$.

Boundaries of cards indicate where the metric coefficient along a
Killing (circle) direction vanishes. Which circle vanishes is
constant over a connected piece of the boundary, even when the
boundary turns a right angle onto a vertical card. Furthermore the
periodicity to eliminate conical singularities is constant along
connected parts of the boundary. For Schwarzschild, the
$\phi$-circle vanishes on both connected boundaries and has
periodicity $2\pi$.

Although not a full replacement for understanding causal
structure, it is interesting to consider geodesic trajectories on
card diagrams. For example when a light ray is incident from a
horizontal card onto a horizon (to enter the upper vertical card),
it must turn and meet that horizon perpendicularly.  It then
appears on the vertical card, again perpendicular to the horizon.
Only those light rays which go from the lower vertical card to the
upper vertical card directly can meet the horizon rod at a
non-right angle; these rays would touch the vertex of the $\times$
in a Schwarzschild Penrose diagram. When a light ray on a vertical
card hits a boundary where a spacelike circle vanishes, it bounces
back at the same angle as drawn on the card relative to the
perpendicular.

Spacetimes with a symmetry group larger than the minimal Weyl
symmetry can have more than one card diagram representation.
Multiple diagrams exist when there is more than one equivalent way
to choose $(D-2)$ Killing congruences on the spacetime manifold.
Examples we explicitly discuss in Sec.~\ref{bubble-sbrane-subsec}
are the 4d Witten bubble and the 4d S-Reissner-Nordstr\o m (S-RN)
which have three card diagrams corresponding to the three types of
Killing congruences on dS$_2$ and ${\bf H}_2$.  These different
card diagrams are associated for example with global, patched, and
Poincar\'e coordinates for dS$_2$ and the different
representations will have different applications and reveal
different information. On the other hand the S-Kerr solution,
whose card diagram is discussed in Sec.~\ref{Kerrsubsec}, has
symmetry group $U(1)\times {\bf R}$ and its unique card
representation looks like the `elliptic' representation of S-RN.

\subsubsection{Our deck of cards:  The building blocks for Weyl spacetimes}

All spacetimes, new and old, in this paper are built from the
following card types.

Horizontal cards are always half-planes.  They may however have
one or more branch cuts which may be taken to run perpendicular to
the $z$-axis. Undoing one branch cut leads to a horizontal strip
with two boundaries; multiple branch cuts lead to some open
subset, with boundary, of a Riemann surface.

Vertical cards may be noncompact: Full planes with or without a
pair of special null lines; half-planes with vertical or
horizontal (horizon) boundary; or quarter planes at any $45^\circ$
orientation.  Vertical cards may also be compact:  Squares with a
pair of special null lines; or 45-45-90 triangles at any
$45^\circ$ orientation.  All horizontal and vertical boundaries
represent where a Killing circle vanishes and hence the end of the
card.  All $45^\circ$ null boundaries represent instances of
${\cal I}^\pm$.

It is satisfying that for a variety of spacetimes including those
in \cite{joneswangfuture,Astefanesei:2005eq}, the cards are always
of the above rigid types.

There is one basic procedure which can be performed on vertical
cards and their corresponding Weyl solutions. It is the analytic
continuation $2\gamma\to 2\gamma+ i \pi$, which is allowed since
$\gamma$ is determined by first order PDEs and $e^{2\gamma}$ is
real of either sign on vertical cards.  This continuation is
equivalent to multiplying the metric by a minus sign and then
analytically continuing the $D-2$ Killing directions.  For charged
generalizations of Weyl solutions (see Appendix B), this procedure
does not affect the reality of the 1-form gauge field. We call
this analytic continuation a $\gamma$-flip since the way it acts
on a card is to geometrically flip it about a $45^\circ$ null line
(for example, look at the vertical cards in Figs.~\ref{S-RN-3} and
\ref{bubblecard3}).

\section{Card diagrams}

In this section we construct the card diagrams for a wide
assortment of solutions including black holes, bubbles and
S-branes. The card diagrams are shown to be useful in representing
continuous changes in the global spacetime structure such as how
Reissner-Nordstr\o m black holes change as we take their
chargeless and extremal limits.  For the superextremal black holes
we discuss how to deal with branch points and cuts on horizontal
cards.  The card diagram also clearly represents the Kerr ring
singularity and how traversing the interior of the ring leads to a
second asymptotic spacetime. The 5d black ring solution,
associated C-metric type solutions and twisted S-branes are also
discussed.

Furthermore analytic continuation has an interesting
interpretation in terms of card diagrams.  We will describe the
effect of analytic continuation on the card diagrams by examining
two known analytic continuations of the Reissner-Nordstr\o m black
hole, the charged bubble and the S0-brane which we also call S-RN.
These time dependent spacetimes each have three card diagram
representation and two are obtained via different analytic
continuations in Weyl coordinates.  The Witten bubble and S-RN are
related to each other by what we call a $\gamma$-flip which is a
geometric realization of analytic continuation.

\subsection{Black holes}

\subsubsection{Subextremal $Q^2<M^2$ Reissner-Nordstr\o m black holes}
\label{RNsubsub}

\begin{figure}[tp]
\begin{minipage}{70mm}
\begin{center}
\includegraphics[width=6.5cm]{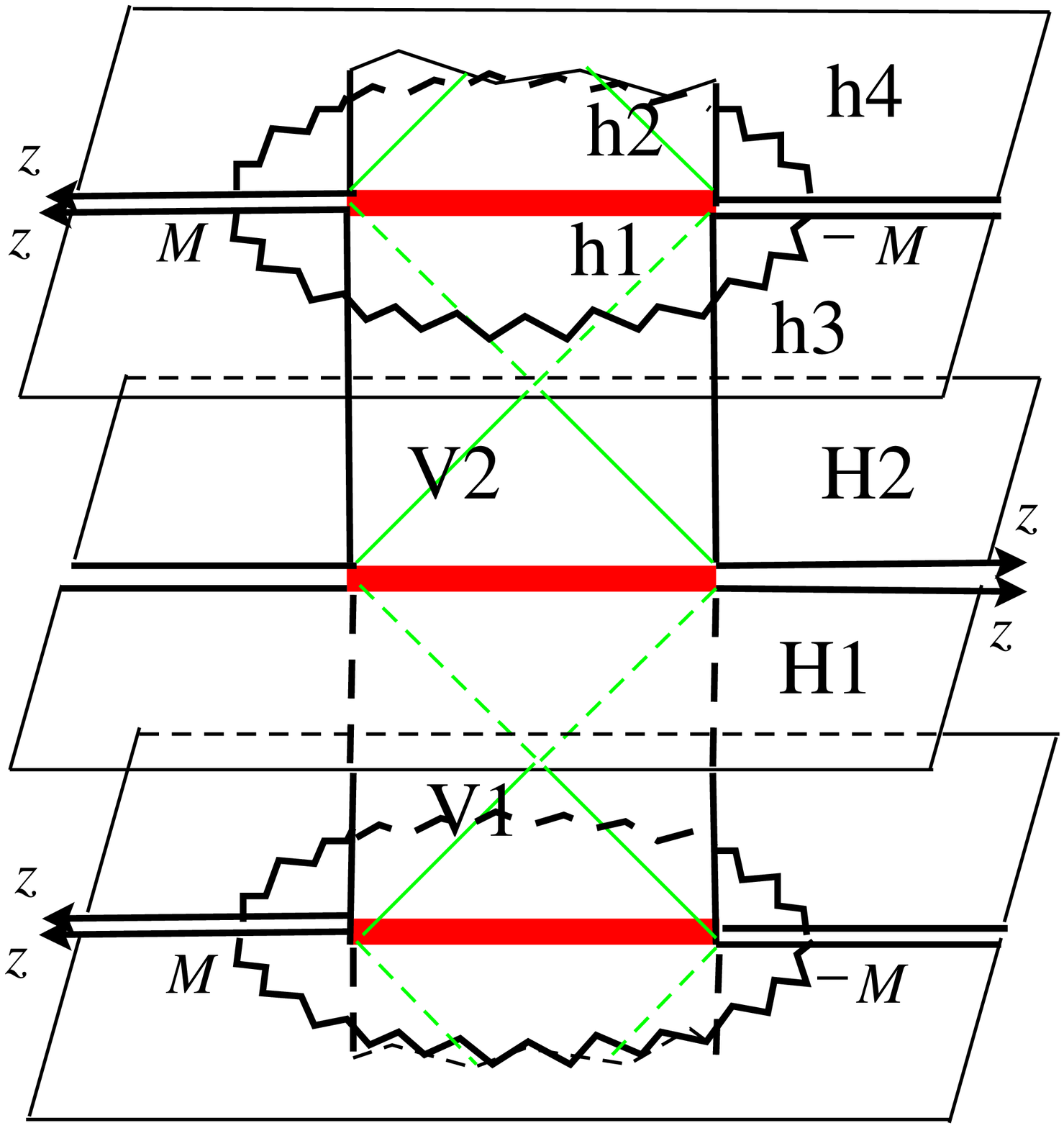}
\caption{The subextremal Reissner-Nordstr\o m card diagram.  The
ellipse singularity has semimajor axes $z=\pm M$ and $\rho=Q$, and
the rod endpoints are the foci on the $z$-axis at
$z=\pm\sqrt{M^2-Q^2}$.} \label{RN-card}
\end{center}
\end{minipage}
\hspace*{15mm}
\begin{minipage}{70mm}
\begin{center}
\includegraphics[width=6cm]{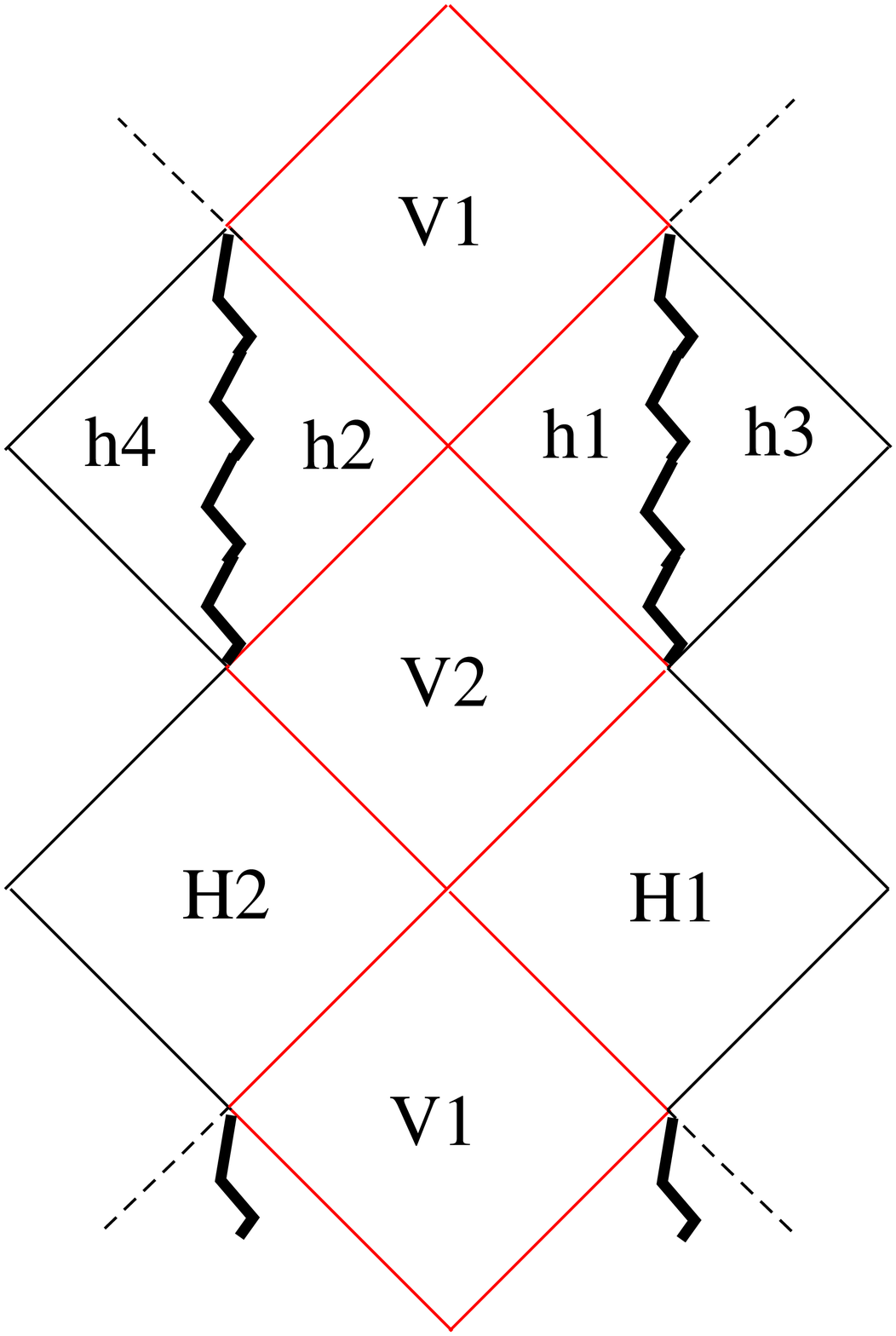}
\caption{The extended Reissner-Nordstr\o m Penrose diagram, including
negative mass universes.}
\label{RN-Penrose}
\end{center}
\end{minipage}
\hspace*{5mm}
\end{figure}

In the usual coordinates the Reissner-Nordstr\o m black hole takes
the form
\begin{eqnarray}
ds^2&=&-\Big(1-{2M\over r}+{Q^2\over r^2}\Big)dt^2+\Big(1-{2M\over
r}+{Q^2\over r^2}\Big)^{-1}dr^2
+r^2(d\theta^2+\sin^2\theta d\phi^2)\label{RNSchwarz}\\
A&=&Qdt/r\nonumber
\end{eqnarray}
Using the coordinate transformation
\begin{equation}
\rho=\sqrt{r^2-2Mr+Q^2}\sin\theta,\qquad
z=(r-M)\cos\theta\label{RNtrans}
\end{equation}
we find the Weyl form of Reissner-Nordstr\o m black hole
\begin{eqnarray}
ds^2&=&-f dt^2+f^{-1}(e^{2\gamma}(d\rho^2+dz^2)+\rho^2 d\phi^2) \label{RNsoln} \label{RNWeyl}\\
f &=& \frac{(R_+ + R_-)^2 - 4(M^2-Q^2)}{(R_+ + R_- +2M)^2} \nonumber \\
e^{2\gamma}&=&\frac{(R_+ + R_-)^2 - 4(M^2-Q^2)}{4 R_+ R_-} \nonumber \\
A&=& -\frac{2Qdt}{R_+ + R_- +2M} \nonumber \\
R_{\pm}&=& \sqrt{\rho^2 + (z\pm \sqrt{M^2-Q^2})^2}= r-M \pm
\sqrt{M^2-Q^2} \cos\theta \nonumber
\end{eqnarray} and the card
diagram for $Q^2<M^2$ is shown in Fig.~\ref{RN-card}.  The
construction of the card diagram proceeds along similar lines to
the Schwarzschild card diagram.  There are two adjacent horizontal
half-planes, H1 and H2, which represent the positive mass
asymptotically flat regions. The outer horizon at
$r_+=M+\sqrt{M^2+Q^2}$ is represented in Weyl space as a rod which
lies along the $z$-axis for $-\sqrt{M^2-Q^2}<z<\sqrt{M^2-Q^2}$.
The vertical cards, V1 and V2, are squares of length
$2\sqrt{M^2-Q^2}$ and the diagonal lines connecting opposite
corners of the square are special null lines. The top of V2 is the
$r_-=M-\sqrt{M^2-Q^2}$ rod which is a four-card inner horizon. The
black hole singularity no longer is on the edge of V2 but instead
is on the outer boundary of the horizontal h1 and h2 regions, at
$\rho^2/Q^2+z^2/M^2=1$. Now the singularity is timelike and
avoidable from the view of an observer on a vertical card. The
rest of those horizontal cards, regions h3 and h4, are $r<0$ or
equivalently $M<0$ nakedly singular RN spacetimes.

At each horizon, the card diagram is continued vertically to
obtain an infinite tower of cards.  In Fig.~\ref{RN-Penrose} we
show the Penrose diagram for comparison.

Although the chargeless $Q\to 0$ limit is hard to understand from
Penrose diagrams, it is easy to understand using the card diagram
in Fig.~\ref{RN-card}; the vertical card expands to a $2M\times
2M$ square and the singularity degenerates to a line segment
coinciding with the inner horizon. Regions h1 and h2 disappear so
the singularity is now `visible' from V1 and V2 as well as h3 and
h4; the singularity is spacelike relative to vertical cards and
timelike for horizontal card observers.  This achieves the
Schwarzschild infinite array of cards in
Fig.~\ref{FullSchwarzCardfig}.

\subsubsection{Extremal $Q^2=M^2$ Reissner-Nordstr\o m black hole}
\begin{figure}[htb]
\begin{center}
\epsfxsize=4in\leavevmode\epsfbox{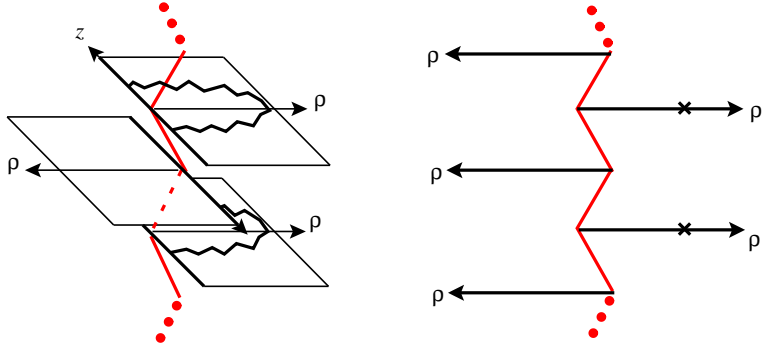} \caption{The
extremal RN card diagram.  The origin of each half-plane is
connected to the origin of its two neighbors.  This is depicted
via a zig-zag line connecting adjacent origins, in analogy to the
Poincar\'e horizons for AdS$_2$. A side-view $z=0$ cross section
is also shown.} \label{RNextremalfig}
\end{center}
\end{figure}

Starting from the above card diagram we now examine the extremal
limit $Q\to \pm M$. In this case the vertical cards which
represent the regions between the two horizons get smaller and
disappear.  When $Q=M$, the horizontal cards are now only attached
at point-like extremal-horizons and only half of the horizontal
cards remain connected, see Fig.~\ref{RNextremalfig}. The region
near the point-horizons are anti-de Sitter throats although cards
themselves cannot adequately depict the throat region. The
throat is a `connected' sequence of points on vertically adjacent
horizontal cards.

To understand how to go across the horizon, it is important to
remember that there are special null lines in $r_-<r<r_+$. In the
extremal case the focal distances are equal ($R=R_+=R_-$) and the
null lines $R_\pm=0$ degenerate to the origin.  So when we pass
through the throat to an adjacent $r<M$ horizontal card, the sign
changes for all occurrences of $R=\sqrt{\rho^2+z^2}=r-M$. Half the
cards are now absent relative to the subextremal case since we no
longer perform analytic continuation to get the vertical card.  We
can only go from positive real values of $\rho$ to negative values
of $\rho$ alternatively.  The singularity appears as a semicircle
on the $r<M$ cards.

For axisymmetric Majumdar-Papapetrou solutions, this `sign change rule'
agrees
with that in \cite{Myers:rx}.  Our analysis also applies to non-MP
axisymmetric solutions such as the dihole
\cite{Bonnor,Chandrasekhar:ds,Emparan:1999au}.

\subsubsection{Superextremal $Q^2>M^2$ Reissner-Nordstr\o m  black holes}

\begin{figure}[htb]
\begin{center}
\epsfxsize=4in\leavevmode\epsfbox{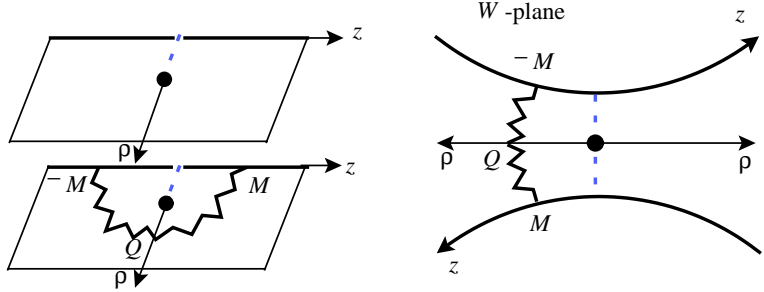} \caption{(a) The
superextremal RN black hole as two half-planes connected along a
dashed branch cut line; (b) after conformal transformation to the
$W$-plane we obtain a branched, static horizontal card, with two
boundaries and one branch point.} \label{superRNfig}
\end{center}
\end{figure}

The superextremal $Q^2>M^2$ Reissner-Nordstr\o m black hole does
not have horizons or vertical cards.  Its card diagram consists of
two horizontal cards, connected along the branch cut
$0\leq\rho\leq\sqrt{Q^2-M^2}$, $z=0$.  One card has a semi-ellipse
singularity passing through the points $(\rho=0,z=\pm M)$ and
$(\rho=Q, z=0)$ (see Fig.~\ref{superRNfig}(a)).

These two horizontal cards are connected in the same sense as a
branched Riemann sheet. By choosing Weyl's canonical coordinates
(meaning $Z=\rho+iz$ with
$-($Coef$\,dt^2)($Coef$\,d\phi^2)=($Re$\,Z)^2$),  the solution is
no longer accurately represented on the horizontal card.  This can
be seen by examining the coordinate transformation (\ref{RNsoln})
from Schwarzschild coordinates to Weyl coordinates.
For fixed $r$ and varying
$\theta$, the coordinates from $M<r<\infty$ cover the Weyl plane
in semi-ellipses which degenerate to the segment $(0\leq\rho\leq
\sqrt{Q^2-M^2}, z=0)$, which serves as a branch cut; and
$\rho=\sqrt{Q^2-M^2}$
is the branch point.  The range
$-\infty<r<M$ again covers the half-plane with $r=0$ forming an
ellipse singularity.  Crossing the branch cut means choosing
the opposite signs for $R_\pm=\sqrt{\rho^2-(z\pm i\sqrt{Q^2-M^2})^2}$,
and indeed we can think of the superextremal `rod' as being
complex-perpendicular
to the Weyl $Z$-plane.

This double cover of the Weyl plane can be fixed
by taking a holomorphic square root; this preserves
the conformally Euclidean character of the card diagram.  By choosing
a new coordinate $W=\sqrt{Z-\sqrt{Q^2-M^2}}$, we map both the
positive and negative-mass universes into the region $({\rm
Im}\,W)^2-({\rm Re}\,W)^2\leq\sqrt{Q^2-M^2}$ (see Fig.
\ref{superRNfig}(b)).  The image of the $z$-axis boundary is a
hyperbola where the $\phi$-circle vanishes.  The origin $W=0$ is
the image of the branch point and the image of the line segment
$(0\leq\rho\leq \sqrt{Q^2+M^2}, z=0)$ is a line connecting the two
hyperbolas and intersecting the origin of the $W$-plane.
The singular nature of $e^{2\gamma}\propto 1/R_+
R_-\propto1/|\Delta Z|\propto 1/|W|^2$ has been fixed by
$e^{2\gamma}dZ d\overline{Z}=4|W|^2 e^{2\gamma}dW d\overline{W}$.
Finally the `black hole' singularity is mapped to a curved segment
stretching from one hyperbola line to the other, to the left of
the branch cut.

Alternatively one can use a Schwarz-Christoffel transformation to
map the two universes onto the strip $|{\rm Im}\,W|\leq W_0$. This
is useful when the horizontal card boundaries are horizons as it
allows for multiple horizontal cards to be placed adjacent to each
other at the horizons (Fig.~\ref{SRN2fig}).   This technique of
fixing a horizontal card with a branch point will also be used in
more complicated geometries such as the hyperbolic representation
of S-RN and multi-rod solutions in four and five dimensions
\cite{Jones:2004pz}.

\subsubsection{Kerr black hole} \label{Kerrsubsec}

Written in Weyl-Papapetrou form, the Kerr black hole is
\begin{eqnarray}
ds^2&=&-f(dt-\omega
d\phi)^2+f^{-1}(e^{2\gamma}(d\rho^2+dz^2)+\rho^2d\phi^2),\\
f&=&\frac{(R_++R_-)^2-4M^2+{a^2\over M^2-a^2}(R_+-R_-)^2}
{(R_++R_-+2M)^2+{a^2\over M^2-a^2}(R_+-R_-)^2}, \nonumber \\
e^{2\gamma}&=&\frac{(R_++R_-)^2-4M^2+{a^2\over
M^2-a^2}(R_+-R_-)^2}
{4R_+R_-}, \nonumber \\
\omega&=&\frac{2aM(M+{R_++R_-\over 2})(1-{(R_+-R_-)^2\over
4(M^2-a^2)})} {{1\over 4}(R_++R_-)^2-M^2+a^2{(R_+-R_-)^2\over
4(M^2-a^2)})},  \nonumber
\end{eqnarray}
where $R_\pm=\sqrt{\rho^2+(z\pm \sqrt{M^2-a^2})^2}=r-M \pm
\sqrt{M^2-a^2} \cos\theta$. The transformation to Boyer-Lindquist
coordinates is $\rho=\sqrt{r^2-2Mr+a^2}\sin\theta$,
$z=(r-M)\cos\theta$.

\begin{figure}[htb]
\begin{center}
\epsfxsize=3.5in\leavevmode\epsfbox{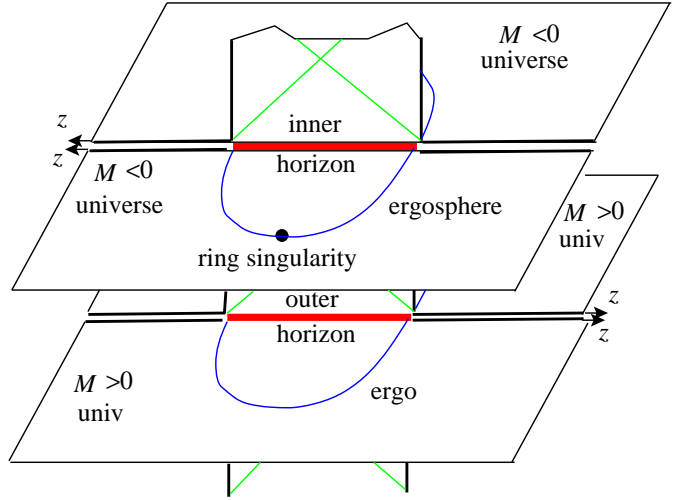} \caption{The
Kerr card diagram.} \label{Kerrcard}
\end{center}
\end{figure}

For $a^2<M^2$ the Kerr black hole card diagram (see
Figure~\ref{Kerrcard}) is similar to Reissner-Nordstr\o m except
that the singularity is now a point and lies at $\rho=a$, $z=0$ on
each horizontal negative-mass card. The outer and inner
ergospheres lie on the positive-and negative-mass cards and are
both described by the curve
$z^2=\alpha^2-(\alpha^2/a^2-1)\rho^2-\rho^4/a^2$ which intersects
the rod endpoints at $z=\pm \alpha=\pm \sqrt{M^2-a^2}$. The
boundary of the region with closed timelike curves is also
described by a quartic polynomial in Weyl coordinates.  Once again
the vertical cards have two special null lines where $R_\pm$
change sign.

The $r=0$ surface in BL coordinates is a semi-ellipse
$\rho^2/a^2+z^2/M^2=1$ on the negative-mass card; but it is not a
distinguished locus on the card diagram.  Attempting to make one
loop around the ring in the Kerr-Schild picture clearly does not
make a closed loop in Weyl space, whereas two loops in the
Kerr-Schild picture will form a single loop around the ring
singularity on the card diagram.  It is also clear that it is
possible to find classical trajectories which avoid the
singularity and which safely escape into a second asymptotically
flat region.

A card diagram for a charged Kerr-Newman solution can
similarly be constructed, with $\alpha=\sqrt{M^2-Q^2-a^2}$.

The extremal Kerr(-Newman) solution has a card diagram like
Fig.~\ref{RNextremalfig}, but the ring singularity is just a point
at $z=0$ and $\rho=M$ on negative-mass cards. Again, $R=R_+=R_-$
and the special null lines degenerate to the origin; crossing the
origin (which is a twisted AdS-type throat) entails changing the
sign of $R$.

The superextremal Kerr(-Newman) solution is similar to the superextremal
RN (Fig.~\ref{superRNfig}(b)) except that the curved-segment singularity
is replaced by a point ($z=0$), and the ergospheres map to an
$\infty$-looking locus centered at $W=0$.

\subsubsection{The black ring}

The 5d static black ring solution of \cite{EmparanWK} is
\begin{eqnarray}\label{blackringsol}
&ds^2&=-{F(x)\over F(y)}dt^2\\
&+&{1\over A^2(x-y)^2}\left[F(x)\Big((y^2-1)d\psi^2 +{F(y)\over
y^2-1}dy^2\Big)+F(y)^2\Big({dx^2\over 1-x^2}+{1-x^2\over
F(x)}d\phi^2\Big)\right]\nonumber
\end{eqnarray}
where $F(x)=1-\mu x$, $F(y)=1-\mu y$, and $0\leq\mu\leq 1$. The
coordinates $x$, $y$ are 4-focus (including $\infty$) or C-metric
\cite{Harmark:2004rm,Emparan:2004wy,bonnorcmetric,bicak}
coordinates that parametrize
a half-plane of Weyl space $\rho\geq0$, $-\infty<z<\infty$:
\begin{eqnarray*}
\rho&=&{1\over A(x-y)^2}\sqrt{F(x)F(y)(1-x^2)(1-y^2)}\\
z&=&{(1-xy)(F(x)-F(y))\over 2A(x-y)^2} \ .
\end{eqnarray*}
The foci are on the $z$-axis at $z=\pm\mu/2A$ and $z=1/2A$. The
black ring horizon is also on the $z$-axis along $-\mu/2A\leq
z\leq \mu/2A$. The $\phi$-circle vanishes along $z\leq-\mu/2A$ and
$\mu/2A\leq z\leq1/2A$, and the $\psi$-circle vanishes along
$z\geq1/2A$.  Curves of constant $y$ degenerate to the
horizon line segment as $y\to -\infty$, and degenerate to the
$(1/2A,\infty)$ ray (better pictured with a conformally
equivalent disk) for $y\to -1$.  Curves of constant $x$ degenerate
to the vanishing $\phi$-circle line segment for $x\to 1$ and to
the ray $(-\infty,-\mu/2A)$ for $x\to -1$.

\begin{figure}[htb]
\begin{center}
\epsfxsize=3.7in\leavevmode\epsfbox{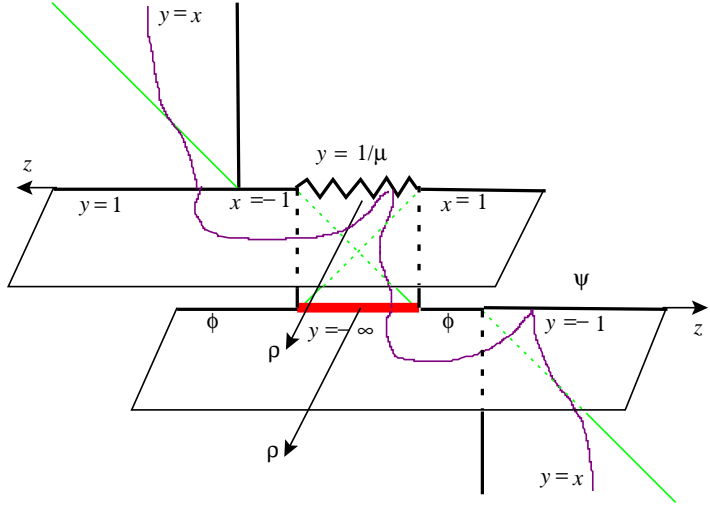} \caption{An
extended card diagram for the black ring, where we have continued
past $y=$\,constant boundaries for $-1\leq x\leq 1$.  A $y$-orbit
is drawn curving through several cards.  Only one of the mirror
pairs of vertical and horizontal cards are drawn at each four card
junction to avoid too many overlapping figures.}
\label{blackringfig}
\end{center}
\end{figure}

The card diagram is easy to construct and is not much different
from the four dimensional Schwarzschild case.  Past $y=-\infty$ we
can go to $y=+\infty$ and hence imaginary $\rho=i\rho'$, and move
up a $\mu/A\times\mu/A$ square with two special null lines.  At
the top of the square, at $y=1/\mu$ we have the curvature
singularity. Continuing again to real $\rho$ and running $y$ down
to $1$, we map out a (negative-mass) horizontal card.  The locus
$y=1$ is the ray $z\geq 1/2A$.  The space closes off here as the
$\psi$-circle vanishes, but we formally continue to illustrate how
C-metric coordinates run on noncompact vertical cards---this is
useful in several applications, such
as the Pleba\'nski-Demia\'nski solution \cite{Plebanski}.
Past $y=1$, we see that for fixed $x$, reducing $y$ down to $x$
makes a topological half-line in a vertical card with a special
null line. Then for $-1<y<x$, we traverse another vertical
card, which we could attach to our original positive-mass
horizontal card along $z>1/2A$ (see Fig. \ref{blackringfig}).

Note that when passing through the black ring horizon at
$y=-\infty$, the Weyl conformal factor \cite{EmparanWK}
$$e^{2\nu}={1+\mu\over 4A}{Y_{23}\over R_1 R_2 R_3}\sqrt{Y_{12}\over
Y_{13}}\sqrt{R_2-\zeta_2\over R_3-\zeta_3},$$ stays real;
$R_3-\zeta_3$ and $Y_{13}$ go negative.  As we pass the special
null lines, explicit appearances of $R_1$ and $R_2$ in the Weyl
functions $e^{2U_i}$, $e^{2\nu}$ change sign.

The charged ring of \cite{Elvang:2003yy} is generated by a
functional transformation and hence inherits a card diagram structure.
In fact, any geometry with $D-2$ Killing directions written in C-metric
coordinates has a card diagram.

\subsection{Charged Witten bubbles and S-branes}
\label{bubble-sbrane-subsec}

The Schwarzschild black hole can be analytically continued to two
different time-dependent geometries, the Witten bubble of
nothing \cite{Witten:1981gj} with a dS$_2$ element and
S-Schwarzschild \cite{ChenYQ,KruczenskiAP,sugraSbranes} with an ${\bf H}_2$
element, and it is instructive to understand how the card diagram changes.

Unlike black hole geometries with their unique card diagrams,
these time dependent geometries can have multiple card diagram
representations. Both the bubble and S-Schwarzschild have three
different card diagram representations corresponding to three
different ways to select Killing congruences. These three types of
Killing congruences can be understood by representing ${\bf H}_2$
as the unit disk (with its conformal infinity being the unit
circle). The orientation-preserving isometries of ${\bf H}_2$ are
those M\"obius transformations preserving the disk, $PSL(2,{\bf
R})$ \cite{Matsuzaki}. M\"obius transformations $z\mapsto
{az+b\over cz+d}$ have two complex fixed points, counted according
to multiplicity.  In the upper half-plane $z=x+i\sigma$
representation, $a$, $b$, $c$, and $d$ are real, so the fixed
points are roots of a real quadratic. Hence they may be (i)
distinct on the real boundary (hyperbolic), (ii) degenerate on the
real boundary (parabolic), or (iii) nonreal complex conjugate
pairs, one interior to the upper half-plane ${\bf H}_2$
(elliptic). Prototypes of Killing fields are (i)
$z\to(1+\epsilon)z$ for the upper half-plane, (ii) $z\to
z+\epsilon$ for the upper half-plane; and (iii) $z\to
e^{i\epsilon}z$ for the disk $|z|<1$. These are the striped,
Poincar\'e, and azimuthal congruences.  In these hyperbolic,
parabolic and elliptic representations, the S-Reissner-Nordstr\o m
(S-RN) and the Witten bubble each have 0, 1, and 2 Weyl foci.

\subsubsection{Elliptic representations and extended card diagrams}
The bubble of nothing in $D$ dimensions has the interpretation as
a semi-classical decay mode of the Kaluza-Klein vacuum.  A spatial
slice is topologically $S^{D-3}\times {\bf R}^2$, where the
${\bf R}^2$ is a cigar with the asymptotic-KK $S^1$ closing at
some fixed $r$, which is an $S^{D-3}$ bubble.
As time passes, the bubble increases in size and `destroys' the spacetime.
The solution is obtained as an analytic continuation
of a black hole and can be generalized to incorporate gauge fields.

\begin{figure}[htb]
\begin{center}
\epsfxsize=4.4in\leavevmode\epsfbox{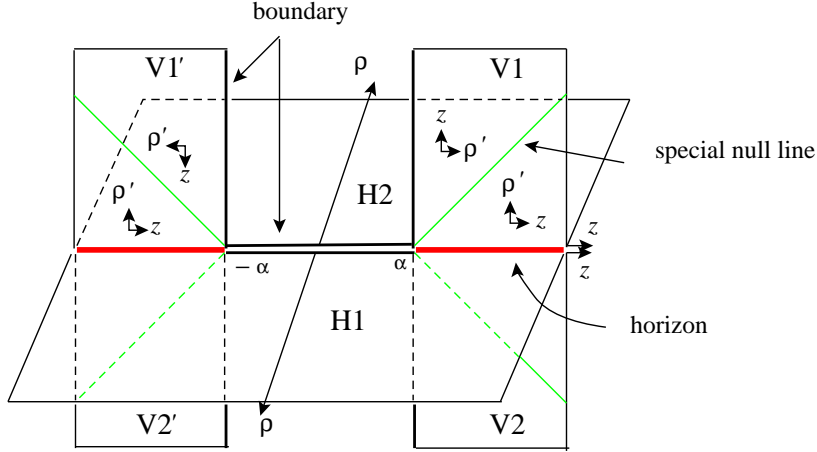} \caption{The
elliptic representation of the non-singular charged Witten bubble
contains two horizons and four special null lines.  Here,
$\alpha=\sqrt{M^2+Q^2}$.} \label{bubblecard1}
\end{center}
\end{figure}

The electrically charged bubble of nothing in its elliptic
representation is gotten from (\ref{RNSchwarz}) by sending $t\to
ix^4$, $\phi\to i\phi$, and to keep the field strength real we
need $Q\to -iQ$. The metric is
\begin{equation}
ds^2=(1-\frac{2M}{r}-\frac{Q^2}{r^2})(dx^4)^2+(1-\frac{2M}{r}-\frac{Q^2}{r^2})^{-1}dr^2
+r^2(d\theta^2-\sin^2\theta d\phi^2) \label{Bubble-metric}
\end{equation}
At $\theta=0,\pi$ there are clearly Rindler-type horizons about
which we analytically continue $\theta$ and obtain the rest of
dS$_2$, $-d\theta^2+\sinh^2\theta d\phi^2$.  These six patches
will precisely correspond to the six cards of
Fig.~\ref{bubblecard1}.

Let us now turn to the effect of the analytic continuation of the
Reissner-Nordstr\o m black hole in Weyl coordinates.   In Weyl
coordinates, the effect of wick rotating $t\to ix^4$ turns the
horizon of the Schwarzschild card into an $x^4$-boundary, while
$\phi\to i\phi$ turns the boundaries on the horizontal card into
noncompact acceleration horizons along the rays $|z|\geq M$,
$\rho=0$.  At the two horizons sending $\theta\to 0 \pm
i\theta,\pi\pm i\theta$ corresponds to $\rho\to \pm i\rho'$ along
the $|z|\geq\sqrt{M^2+Q^2}$ rays. We find vertical noncompact one
eighth-plane cards with special null lines along $\rho'= z-M$ for
$z\geq M$ and $\rho'=-M-z$ for $z\leq -M$.  Each piece of the
plane is a doubly covered triangle and it is necessary to change
branches of the function $R_\pm$ at the null line, as they appear
in (\ref{RNWeyl}) to transform the card into a single covered
quarter plane card.

The elliptic (and as we will shortly see the hyperbolic)
representations of the Witten bubble are simply obtained because
their dS$_2$ Killing congruences are trivially obtained from those
on $S^2$.  Specifically, take the $S^2$ embedding into flat space
with $Z=\cos\theta$, $X=\sin\theta\cos\phi$,
$Y=\sin\theta\sin\phi$.  Sending $\phi\to i\phi$ has the effect
$Y\to iY'$ so the surface becomes $X^2-Y'^2+Z^2=1$ embedded in
Minkowski space, or dS$_2$ with $\phi$ as an elliptic (azimuthal)
congruence. On the other hand sending $\theta\to \pi/2+i\theta$,
has the effect $Z\to iZ'$ giving $X^2+Y^2-Z'^2=1$, which is dS$_2$
again but with $\phi$ as a hyperbolic (striped) congruence.

Next the S-brane solution of
\cite{ChenYQ,KruczenskiAP,sugraSbranes}
\begin{equation}
ds^2=(1+\frac{2M}{t}-\frac{Q^2}{t^2})(dx^4)^2-(1+\frac{2M}{t}-\frac{Q^2}{t^2})^{-1}dt^2
+t^2(d\theta^2+\sinh^2\theta d\phi^2) \label{S-Sch-metric}
\end{equation}
can also be gotten from (\ref{RNSchwarz}) by taking $t\to ix^4$,
$\theta\to i\theta$, $r\to it$, and $M\to i M$.  From
(\ref{RNtrans}) we see that in Weyl's coordinates this analytic
continuation of RN can be implemented by sending $t\to ix^4$,
$z\to i\tau$, $M\to i M$, up to a real coordinate transformation.

\begin{figure}[htb]
\begin{center}
\epsfxsize=4in\leavevmode\epsfbox{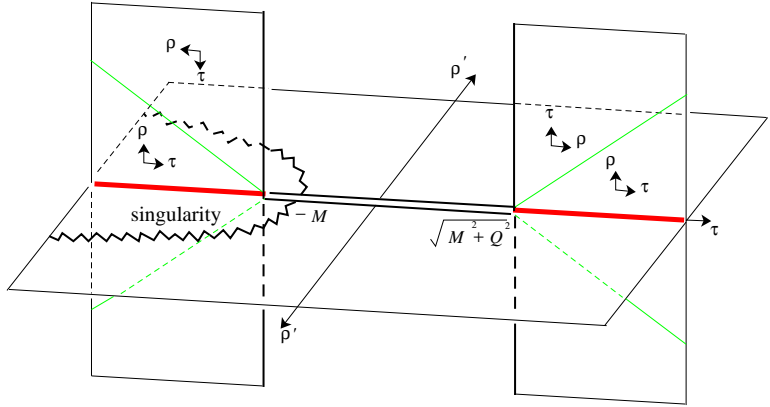}
\caption{The elliptic S-RN card diagram is similar to
Fig.~\ref{bubblecard1} except there is a singularity in the shape
of a hyperbola which does not intersect the horizons.}
\label{chargedSbranecard}
\end{center}
\end{figure}

\begin{figure}[htb]
\begin{center}
\epsfxsize=3in\leavevmode\epsfbox{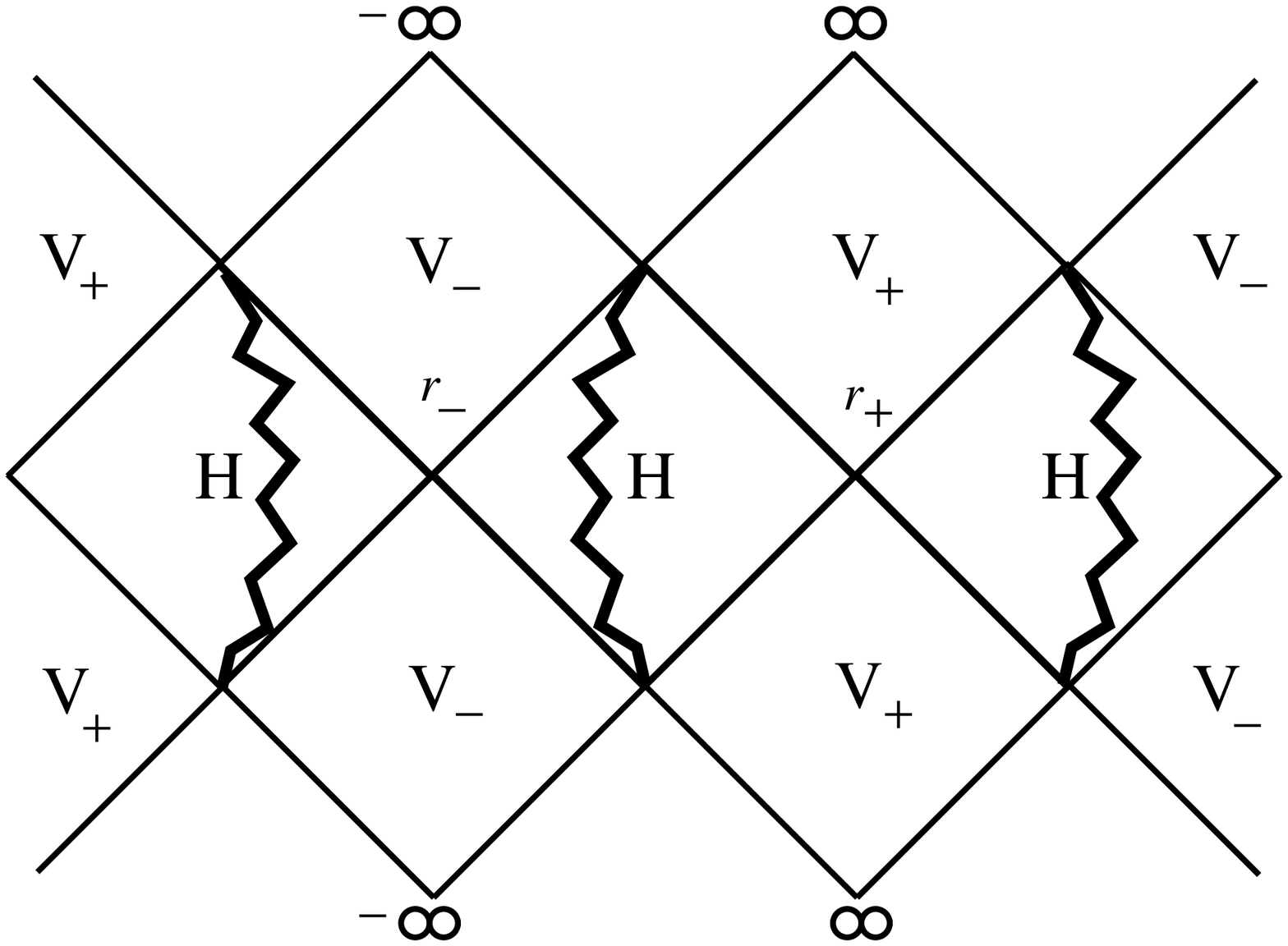} \caption{The
Penrose diagram for S-RN.} \label{SRNPenrosefig}
\end{center}
\end{figure}

The card diagram for elliptic S-RN (Fig.~\ref{chargedSbranecard})
has the same structure as the Witten bubble
(Fig.~\ref{bubblecard1}) except that the 6-segment boundary is now
$\theta=0$ where the $\phi$-circle vanishes.  The right and left
horizons are at $t=t_\pm=M\pm\sqrt{M^2+Q^2}$. The $t=0$
singularity is a hyperbola on the $t_-\leq t\leq t_+$ horizontal
cards parametrized as
$(\rho',\tau)=(|Q|\sinh\theta,-M\cosh\theta)$. Any $Q\neq0$ gives
the same qualitative diagram.  The `smaller' connected universe on
the card diagram is the negative-mass version of S-RN. Either sign
of the mass gives a universe that is cosmologically singular.  The
Penrose diagram is given in Fig.~\ref{SRNPenrosefig}.

In the limit $Q\to 0$, the hyperbola singularity degenerates to a
straight line covering the horizon at $t=0$.  The two $t>2M$
vertical cards and two $0<t<2M$ horizontal cards then form a
positive-mass S-Schwarzschild, while each $t<2M$ card forms a
negative-mass S-Schwarzschild whose singularities begin or end the
spacetime.

One can alternatively form the elliptic S-Schwarzschild from the
elliptic Witten bubble by performing the $\gamma$-flip on any
vertical card.  This procedure is immediate; the net continuation
from Schwarzschild is $\theta\to i\theta$, $g_{\mu\nu}\to
-g_{\mu\nu}$, and avoids $r\to it$, $z\to i\tau$, and $M\to iM$.

Note how the card representation of the S-brane is quite different
from the black hole card diagram while the Penrose diagrams of the
two spacetimes are nearly just related by ninety degree rotation.
This is because the card diagram shows the compact or noncompact
$\theta$ direction.

\begin{figure}[htb]
\begin{center}
\epsfxsize=3.5in\leavevmode\epsfbox{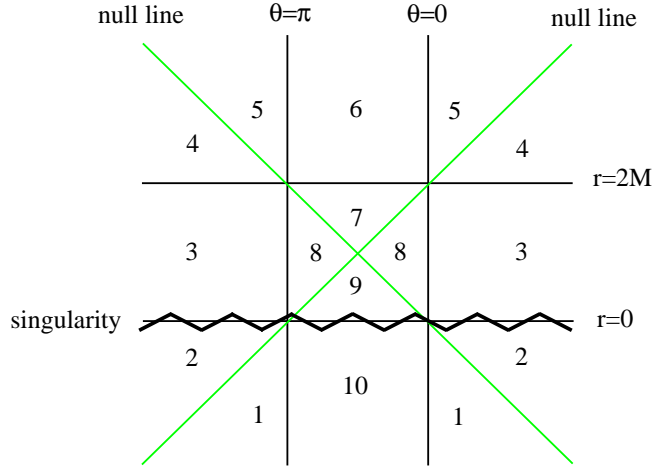} \caption{The
Schwarzschild black hole, Witten bubble and S-Schwarzschild are
all related by $\gamma$-flips and so their (elliptic) card
diagrams may be combined into this extended diagram.}
\label{fullsSchwarz}
\end{center}
\end{figure}

The elliptic form of the card diagrams show that Schwarzschild
S-brane, Witten bubble and Schwarzschild solutions have similar
structures and in fact they are all related by $\gamma$-flips and
trivial Killing continuations.\footnote{Perturbed solutions that
can only be obtained from $z\to i\tau$ are considered to be less
trivial.} Solutions which are related in this manner may be
conveniently drawn together in one diagram which simultaneously
displays all of their card diagrams.  For example in
Fig.~\ref{fullsSchwarz} the S-Schwarzschild solution comprises
regions $1,2,3,4,5$, the Witten bubble regions $4,5,6$, and the
Schwarzschild black hole $6,7,8,9,10$. Regions $1,2,10$ correspond
to a singular Witten bubble of negative `mass.'  In this diagram
we also see that the unification of the special null lines as they
extend through all three solutions; we can say $R_+=0$ is the long
$\diagup$-null line and $R_-=0$ is the long $\diagdown$-null line.

The charged Reissner-Nordstr\o m BH/bubble/S-brane solutions
cannot be depicted together on such a diagram because $Q\to iQ$
changes $0<r_-<r_+$ to $r_-<0<r_+$. Similar diagrams can be found
in \cite{Astefanesei:2005eq,bicak}.

\subsubsection{Hyperbolic representations and branch points}
\label{Witten2sec}

The charged Witten bubble
\begin{equation}
ds^2=\Big(1-{2M\over r}-{Q^2\over
r^2}\Big)(dx^4)^2+\Big(1-{2M\over r}-{Q^2\over r^2}\Big)^{-1}dr^2
+r^2(-d\theta^2+\cosh^2\theta d\phi^2)
\end{equation}
can alternatively be obtained from the RN black hole
(\ref{RNSchwarz}) by taking $\theta\to\pi/2+i\theta$ and $t\to
ix^4$, $Q\to -iQ$.  Here, $\theta$ plays the role of time and
$\theta=0$ is the time where the bubble `has minimum size.'  (This
statement has meaning if we break $SO(2,1)$ symmetry.)  To achieve
this in Weyl's coordinates, we put $z\to i\tau$, $t\to ix^4$,
$Q\to -iQ$; the resulting space is equivalent to Witten's bubble
by the real coordinate transformation
\begin{equation}
\rho=\sqrt{r^2-2Mr-Q^2}\,\cosh\theta,\qquad \tau=(r-M)\sinh\theta.
\end{equation}
Thus in Weyl coordinates the only difference between the
hyperbolic Witten bubble and the elliptic S-RN is putting $M\to
iM$. Witten's bubble universe is represented in Weyl coordinates
as a vertical half-plane card, $\rho\geq 0$,
$-\infty<\tau<\infty$, where now the $x^4$-circle, and not the
$\phi$-circle, vanishes at $\rho=0$.  This boundary is where the
bubble of nothing begins.  Note that the vertical card now has
Minkowski signature and is conformal to $-d\tau^2+d\rho^2$. This
vertical card does not have special null lines since the foci are
at imaginary values $\tau=\pm iM$, and so the spacetime is covered
only once by the Schwarzschild coordinates. The bubble does have a
rod which is along the imaginary $\tau$ axis and which intersects
the card at the $\rho=0$, $\tau=0$ origin.  The hyperbolic
representation of the charged Witten bubble is therefore just a
vertical half-plane.  In such a case the card diagrams have just
as much causal information as a Penrose diagram and the boundary
of the card easily describes the edge of the bubble of nothing.

We obtain a hyperbolic representation for S-RN from
(\ref{RNSchwarz}) by sending $t\to i x^4$, $M\to i M$, $r\to i r$,
$\theta\to \pi/2+i\theta$, and $\phi\to i\phi$. The fibered
directions are now hyperbolic $d{\bf
H}_2^2=d\theta^2+\cosh^2\theta d\phi^2$. In Weyl coordinates, to
maintain reality of the solution, we begin with RN (\ref{RNWeyl})
and send $t\to ix^4$, $\phi\to i\phi$, $M\to iM$, and must
explicitly change the branch of the square root introducing the
minus sign $R_-\to -R_-$.\footnote{The naturalness of this sign
change is explained in great detail in \cite{Jones:2004pz}.} This
has the interpretation of staying on the same horizontal card and
rotating the rod in the complex $z$-plane.  The foci are then at
($z=\pm i\sqrt{M^2+Q^2}$, $\rho=0$) with their special null lines
intersecting the real half-plane at $z=0$, $\rho=\sqrt{M^2+Q^2}$.
The half-plane is doubly covered, and we will take
$0\leq\rho\leq\sqrt{M^2+Q^2}$ as the branch cut. The sign change
of $R_-$ has effectively reversed the roles of $r$ and $\theta$ so
that, after undoing the branch cut with say a square-root
conformal transformation, $r=r_+>0$ is one boundary-horizon and
$r=r_-<0$ is the other. The hyperbolic angle $\theta$ is unbounded
$-\infty<\theta<\infty$.

\begin{figure}[htb]
\begin{center}
\epsfxsize=2.5in\leavevmode\epsfbox{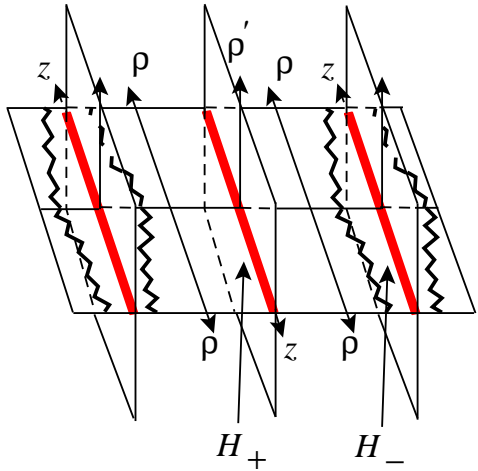} \caption{Hyperbolic
S-RN card diagram after transformation to the $W$-plane on the
horizontal cards is without boundaries.  The singularities are
closer to the ${\cal H}_-$ horizon than to ${\cal
H}_+$.}\label{SRN2fig}
\end{center}
\end{figure}

The doubly-covered half-plane is physically cut into two by the
$r=0$ singularity. At each horizon $r=r_\pm$ we have a four-card
junction; the double half-plane horizontal card meets another
mirror copy as well as two vertical cards.  The vertical cards are
at $\rho\to\pm i\rho'$ and $-\infty<z<\infty$ and have no special
null lines or other features.  A full card diagram is shown in
Fig.~\ref{SRN2fig}. Note that there are no boundaries of this card
diagram where a spacelike Killing circle vanishes.

Setting $R_\pm=\sqrt{\rho^2+(z\pm i\sqrt{M^2+Q^2})^2}$ and
$R=R_+$, the explicit form of hyperbolic S-RN on the horizontal
card is
\begin{eqnarray}\label{srn2}
ds^2&=&-f(dx^4)^2+f^{-1}(e^{2\gamma}(d\rho^2+dz^2)+\rho^2d\phi^2),\\
f&=&{M^2+Q^2-({\rm Im}\,R)^2\over ({\rm Im}\,R+M)^2},\nonumber\\
e^{2\gamma}&=&{M^2+Q^2-({\rm Im}\,R)^2\over |R|^2},\nonumber\\
A&=&{Q\,dx^4\over {\rm Im}\,R+M}.\nonumber
\end{eqnarray}

We can arrive at this spacetime in a simpler way.  Take the RN
black hole and analytically continue to get the hyperbolic charged
Witten bubbles. These universes are nonsingular for $r\geq r_+$ or
$r\leq r_-$. They have boundaries at $r=r_\pm$ where the
$x^4$-circle vanishes. Now, turn these universes on their sides
with the $\gamma$-flip. This allows us to decompactify $x^4$ and
$r=r_\pm$ are now Milne horizons---we are looking precisely at the
vertical cards of Fig.~\ref{SRN2fig}, and they are connected in a
card diagram by an $r_-\leq r\leq r_+$ card which is now
accessible. We see that generally, vertical half-plane cards
parametrized in spherical prolate fashion with no special null
lines, when turned on their sides, connect to branched horizontal
cards.

The $Q\to 0$ limit (hyperbolic S-Schwarzschild) of the card
diagram is easy to picture:  The singularities of
Fig.~\ref{SRN2fig} collapse onto the $r=r_-$ horizon.

One may wonder what happens if we take $z\to i\tau$, on the
horizontal card for hyperbolic S-RN, and achieves a vertical card
sandwiched between special null lines at
$\rho=\sqrt{M^2+Q^2}+|\tau|$.  This must unfold to give a vertical
plane with two intersecting special null lines.  This card diagram
structure is discussed in \cite{Jones:2004pz,joneswangfuture}.

\subsubsection{Parabolic representations}
\label{thirdcard-Poincare}

\begin{figure}[tp]
\hspace*{5mm}
\begin{minipage}{70mm}
\begin{center}
\includegraphics[width=8cm]{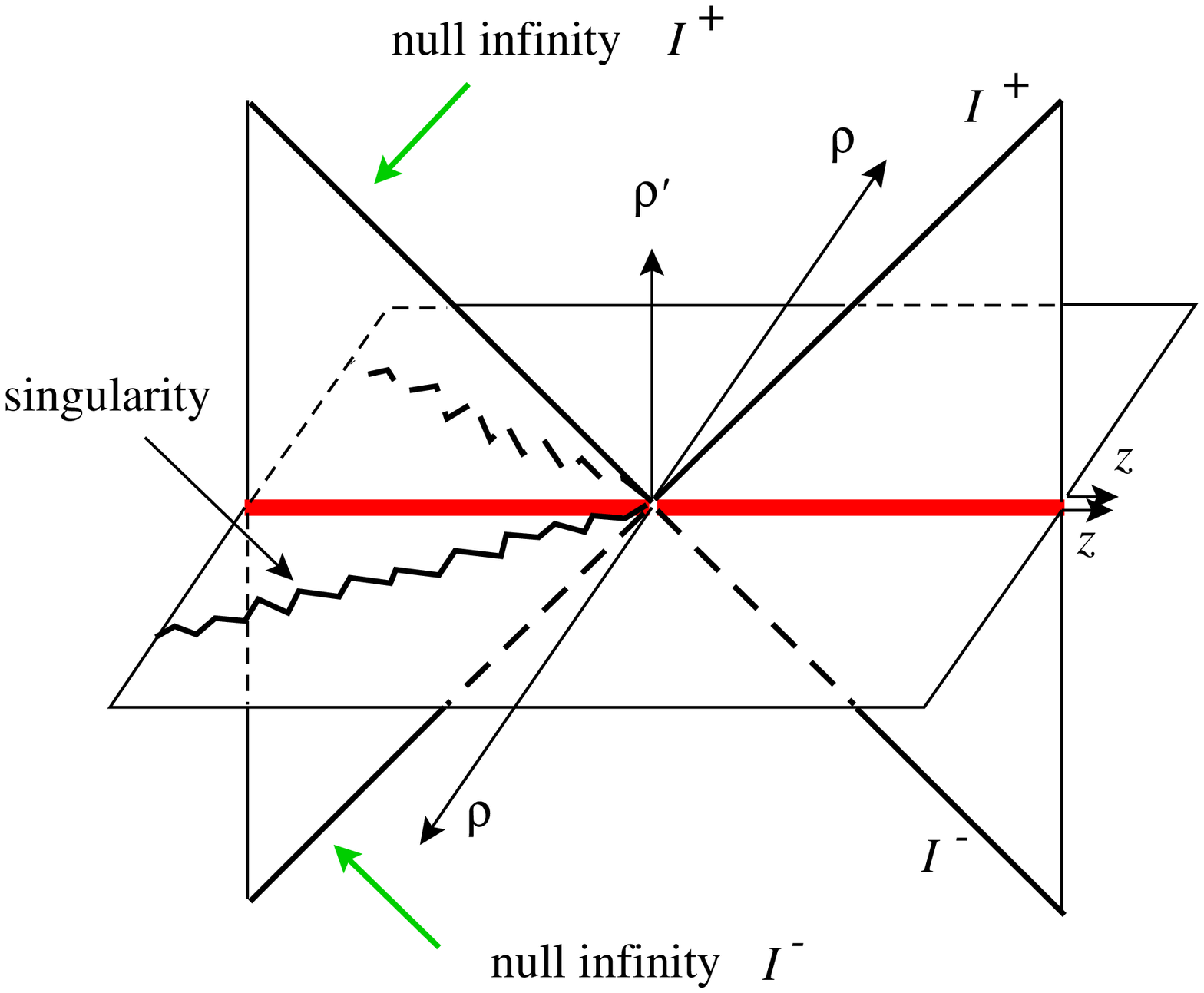}
\caption{The parabolic card diagram representation for
S-Reissner-Nordstr\o m.  The $45^\circ$ vertical lines represent null
infinity.} \label{S-RN-3}
\end{center}
\end{minipage}
\hspace*{15mm}
\begin{minipage}{70mm}
\begin{center}
\includegraphics[width=7cm]{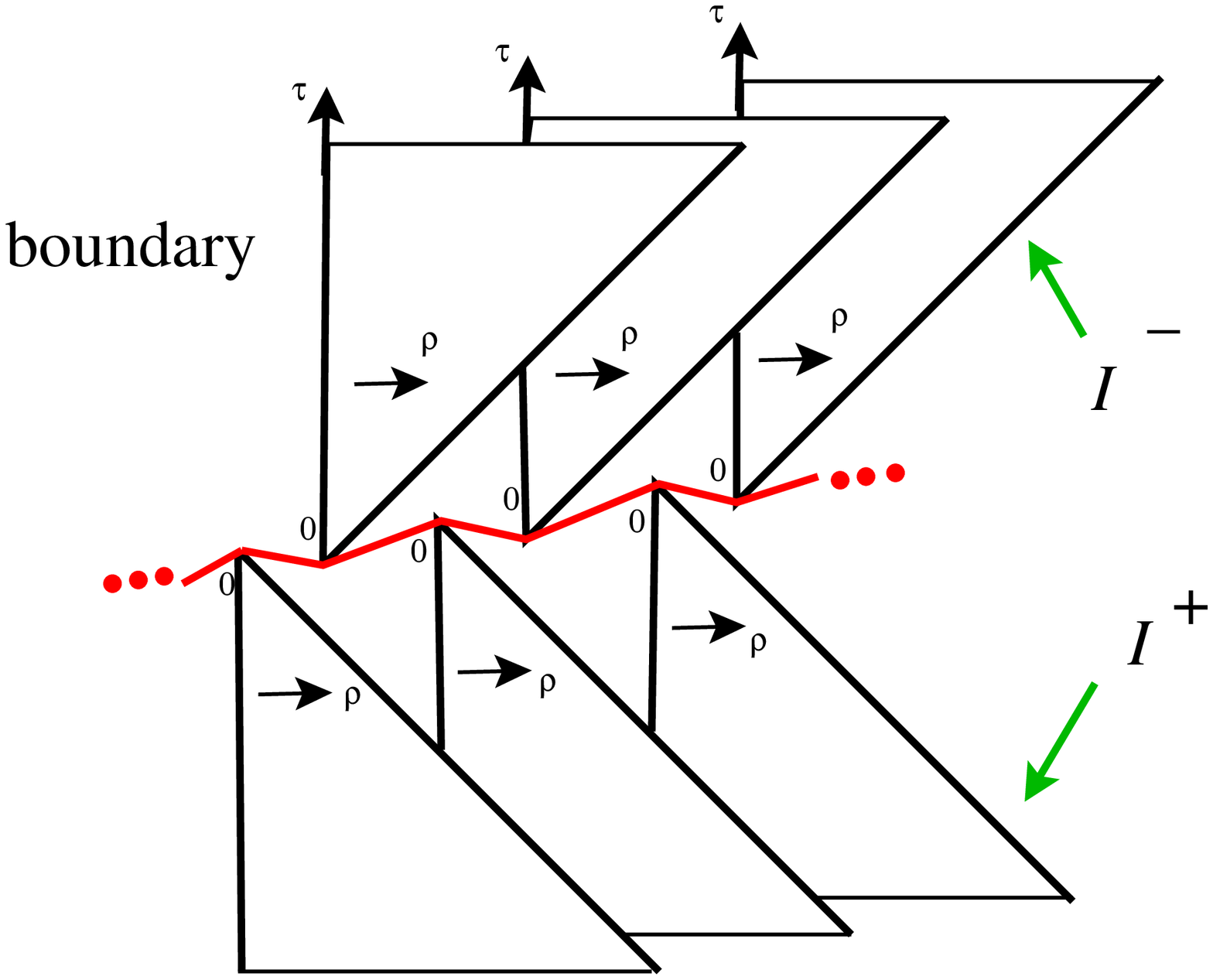}
\caption{The parabolic representation of the Witten bubble
contains an infinite number of $45^\circ$ wedge vertical cards
pointing up and down with each wedge joined to two others at the
tip in a dS$_2$ fashion.} \label{bubblecard3}
\end{center}
\end{minipage}
\hspace*{5mm}
\end{figure}

There is a third way to put a Killing congruence on ${\bf H}_2$ or
dS$_2$ using parabolic or Poincar\'e coordinates.  Parametrizing
hyperbolic space (Euclideanized AdS$_2$) as $ds^2=\sigma^2
dx^2+{d\sigma^2\over \sigma^2}$, and keeping the Schwarzschild
S-brane coordinate $r$ and the usual $x^4$ we get a Poincar\'e
Weyl representation of S-RN spacetime
\cite{Gutperle:2002ai,ChenYQ,KruczenskiAP}.  It is
\begin{eqnarray}
ds^2&=&f(dx^4)^2-f^{-1}(e^{2\gamma}(-d\rho'^2+dz^2)+\rho'^2dx^2),\\
A&=&Qdx^4/r, \nonumber\\
f&=&(1-2M/r-Q^2/r^2),\nonumber\\
e^{2\gamma}&=&{r^2-2Mr-Q^2\over \sigma^2(M^2+Q^2)},\nonumber\\
\rho'&=&\sigma\sqrt{r^2-2Mr-Q^2},\nonumber\\
z&=&\sigma(r-M).\nonumber
\end{eqnarray}
In this Weyl representation, $\rho'$ is timelike on $r\geq r_+$
vertical cards which are noncompact $45^\circ$ wedges,
$0\leq\pm\rho'<z$. This connects along $z>0$ to an $r_-\leq r\leq
r_+$ horizontal card; $r\leq r_-$ vertical cards attaches to
$z<0$. So this is similar to the elliptic representation of S-RN,
except the line segment $-\sqrt{M^2+Q^2}<z<\sqrt{M^2+Q^2}$ has
collapsed and the special null lines are now conformal null
infinity (Fig.~\ref{S-RN-3}). The singularity on the horizontal
card is particularly easy to describe in these coordinates.  On
the first horizontal card it is on a ray $z/\rho=-M/Q$ and on the
second card it is on a ray $z/\rho=M/Q$.

If we take the $r\geq r_+$ (or $r\leq r_-$) $45^\circ$ wedges and
turn them on their sides via the $\gamma$-flip, we get the
parabolic version of the $r\geq r_+$ (or $r\leq r_-$) charged
Witten bubble. The line which used to be the horizon in the
S-brane card diagram becomes a boundary which is the `minimum
volume' sphere, at $\rho=0$. Time is now purely along the $\tau$
direction as in the hyperbolic Witten bubble.  The special null
line is still ${\cal I}^\pm$ since $\rho=|\tau|$ corresponds to
$r\to\infty$.  The vertex of the triangular card is not the end of
the spacetime.  These wedge cards only represent $\sigma>0$ and so
the card diagram should be extended to $\sigma<0$.  The card
diagram is an infinite array of $45^\circ$ wedge cards pointing up
and an infinite number pointing down.  The vertex of each upward
card is attached to its nearest two downward neighbors (one to the
left and one to the right), in the dS$_2$ fashion as shown in
Figure~\ref{bubblecard3}. One can identify cards so only needs one
upward and one downward card with two attachments. Although this
card diagram is not the most obvious representation of the Witten
bubble, it is useful in understanding the S-Kerr solution of the
next subsection as well as the more complicated S-dihole ${\cal
U}$, ${\cal U}_\pm$ universes of \cite{Jones:2004pz}.

\subsection{S-Kerr}

\begin{figure}[htb]
\begin{center}
\epsfxsize=2.5in\leavevmode\epsfbox{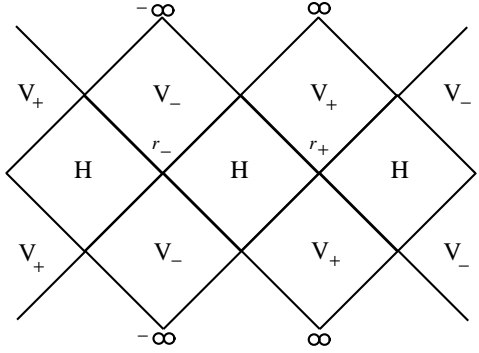}
\caption{Subextremal $a^2<M^2 $ S-Kerr Penrose diagram.  V$_\pm$
map to vertical cards of positive and negative `mass,' while all H
diamonds give identical horizontal cards.  It is possible to
identify cards (say, every other H diamond) so there are only
a finite number of regions
in the spacetime.} \label{SKerrPenrosefig}
\end{center}
\end{figure}

The twisted S-brane~\cite{Wang:2004by}, see also
\cite{Tasinato:2004dy}, is also known as S-Kerr, and it is another
example of a nonsingular time-dependent solution.  Twisted
S-branes describe the decay of unstable massive strings.  They can
be obtained from the Kerr black hole using the following card
diagram method. Double Killing continue $t\to ix^4$, $\phi\to
i\phi$ to achieve a ${\cal K}_+$ bubble solution, go to the
vertical card via $\theta\to i\theta$, then perform a
$\gamma$-flip to achieve S-Kerr. S-Kerr has symmetry group
$U(1)\times{\bf R}$ and therefore has a unique card diagram. For
the parameter range $a^2<M^2$ there are horizons and the card
diagram structure is that of the elliptic bubble
(Fig.~\ref{bubblecard1}). The foci are at $z=\pm\sqrt{M^2-a^2}$.
There is an ergosphere and CTC region on the horizontal card and
has the same qualitative shape as it does for the Kerr black hole
diagram.   In comparison, the Penrose diagram showing (variably
twisted) $x^4$ and the Boyer-Lindquist coordinate $r$ is shown in
Fig.~\ref{SKerrPenrosefig}.

\begin{figure}[htb]
\begin{center}
\epsfxsize=3in\leavevmode\epsfbox{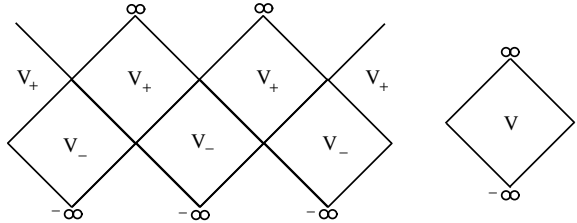}
\caption{The extremal $a^2=M^2$ (left) and superextremal $a^2>M^2$ (right)
S-Kerr Penrose diagrams.} \label{SKerrExtremalPenrosefig}
\end{center}
\end{figure}

In the extremal limit $\pm a\to M$, $\theta$-orbits on the
vertical card shift up relative to the special null line, and any
fixed ($r$, $\theta$) point is sent above the null line.  The
region below the null line disappears in this limit and the
horizontal card collapses to a point. Furthermore, those geodesics
in the upper-right card can only reach the lower-left card (and
the same with upper-left and lower-right), splitting the universe
into connected $45^\circ$ wedges just like the parabolic
representation for the Witten bubble (Fig. \ref{bubblecard3}),
where the connections are  dS$_2$-like twisted throats.

The case $a^2>M^2$ for S-Kerr does not have horizons and can be
represented as a single vertical half-plane card with $\rho\geq 0$
and with no special null lines.  Just like the hyperbolic Witten
bubble, superextremal S-Kerr can be turned on its side with the
$\gamma$-flip, to yield a new spacetime \cite{joneswangfuture}.
In the limit where $|a|\to\infty$, the $\theta$-orbits
flatten out on the card, and the solution becomes flat space.

The Penrose diagrams for extremal and superextremal S-Kerr are
given in Fig.~\ref{SKerrExtremalPenrosefig}.

\section{Discussion}

In this paper we have examined and extended the utility of the
Weyl Ansatz through the construction of an associated Weyl card
diagram.  Weyl coordinates are an excellent choice for clearly
understanding geometric features and so drawing the card diagram
is useful.  The card diagram conveniently captures most of the
interesting properties of a spacetime including its singularities,
horizons, and some aspects of its causal structure and null
infinity.  Card diagrams for families of solutions such as charged
and rotating black holes (and bubbles and S-branes) share similar
features and can change continuously.  They are useful in keeping
track of various analytic continuations and mentally partitioning
complicated spacetimes into simpler regions. The only technical
issues that seem to arise and which we resolved dealt with
branches of Weyl distance square-root functions, special null
lines on vertical cards, and branched horizontal cards.

Here we summarize the solutions in this paper. The card diagrams
correctly capture the different regions of the charged
Reissner-Nordstr\o m black hole and its various charged and
chargeless limits, and its negative mass complement. We also
analyzed the Kerr black hole and its singularity structure.  In
particular the safe passage through the interior of the ring to
the second asymptotic universe through $r=0$ is clearly depicted.

The Witten bubbles and S-branes each have three card diagrams
representations corresponding to the three different choices of
Killing congruences on dS$_2$ or ${\bf H}_2$. The elliptic
representations had two foci and six cards. S-RN had a
cosmological singularity on the horizontal card splitting the card
diagram into two connected universes, whereas the charged bubbles
have the same card structure but are nonsingular. The hyperbolic
representations have no foci: the bubble is a simple vertical
half-plane card and hyperbolic S-RN has a branched horizontal card
which we fixed by a conformal mapping of the half-plane.  Finally
the parabolic representation of the bubble was an infinite array
of $45^\circ$ triangle wedges connected pointwise while S-RN had a
6-card butterfly shape.  This parabolic representation showcased
special null lines serving as null infinity.

S-Reissner Nordstr\o m can be obtained from the bubble in two
ways. One may start with the bubble and analytically continue
$M\to i M$ in Weyl coordinates.  In this case the
hyperbolic/elliptic representation of the bubble maps to the
elliptic/hyperbolic representation of the S-brane.  We also found
that these two solutions can be related by what we called the
$\gamma$-flip, which is conveniently visualized as a flip of the
associated cards about a null line. This procedure maintains the
number of Weyl foci on the card and so maps the
elliptic/parabolic/hyperbolic bubble to the
elliptic/parabolic/hyperbolic S-brane.  The $\gamma$-flip provides
a simple and geometric way to relate Schwarzschild with its two
analytic continuations, the bubble of nothing and the S-brane.  In
fact spacetimes related by $\gamma$-flips can be simultaneously
drawn together in a complexified $r\theta$ (or $\rho,z$) spacetime
diagram.

Just as we used the $\gamma$-flip to turn the hyperbolic Witten
bubble on its side and got hyperbolic S-RN, we can take the
vertical half-plane card diagrams for the Kerr bubble, dihole
wave, superextremal S-Kerr and superextremal S-dihole and apply
the $\gamma$-flip to yield new spacetimes. These new solutions
will be described in \cite{joneswangfuture}.

The card diagram formalism can be further generalized; the
recently developed Weyl-Papapetrou formalism \cite{Harmark:2004rm}
for $D\geq 5$ will yield card diagrams and 5d Kerr-related
solutions will appear in \cite{joneswangfuture}. Furthermore card
diagrams do not require Weyl's canonical coordinates. Spacetimes
with Weyl-type symmetry and yet where Weyl's procedure fails
algebraically can still admit card diagrams.  An example is the
inclusion of a nonzero cosmological constant $\Lambda$, where a
$\gamma$-flip changes the sign of $\Lambda$.  Card diagrams for
pure (A)dS$_D$ space for $D=4,5$ are presented in
\cite{Astefanesei:2005eq}.  Constant-curvature black holes
obtained by quotienting \cite{Cai:2002mr} will also have card
diagrams.  We hope that our methods, and their further
generalizations, will have even greater applicability than to the
multitude of spacetimes already discussed herein.

\section*{Acknowledgements}
We thank D.~Jatkar, A.~Maloney, W.~G.~Ritter, A.~Strominger,
T.~Wiseman and X.~Yin for valuable discussions and comments.
G.~C.~J.~would like to thank the NSF for funding.  J.~E.~W. is
supported in part by the National Science Council, the Center for
Theoretical Physics at National Taiwan University, the National
Center for Theoretical Sciences and would like to thank the
organizers of Strings 2004 for support and a wonderful conference
where part of this research was conducted.


\appendix
\section{Perturbed bubbles and S-branes}

The chargeless Schwarzschild black hole is easily perturbed as
a Weyl solution by adding more rod-horizons, to form Israel-Khan
arrays.  We can use these solutions to smoothly perturb the Witten
bubble in two different ways and also the S-Schwarzschild in two
different ways.  Addition of charge can be done via Weyl's electrification
method \cite{Jones:2004pz,Weylpaper,Fairhurst:2000xh}.

The analytic continuation in Weyl space to obtain the hyperbolic
Witten bubble is precisely the same as in \cite{Jones:2004rg},
except here the Schwarzschild rod crosses $z=0$. Even-in-$z$
Israel-Khan arrays where no rod crosses $z=0$ can be analytically
continued to gravitational wave solutions sourced by imaginary
black holes, ie rods at imaginary time. We can thus generalize the
Witten bubble by adding additional waves by symmetrically placing
more rods in addition to the one which crosses $z=0$. We dub such
an array the `hyperbolic-perturbed Witten bubble.' As these
additional rods are made to cover more of the $z$-axis and are
brought closer and closer to the principal rod, the deformed
Witten bubble solution hangs longer with a minimum-radius
$\phi$-circle.  In the limit where rods occupy the entire
$z$-axis, we get a static flat solution, which is Minkowski
3-space times a fixed-circumference $\phi$-circle.

The hyperbolic-perturbed Witten bubble can be turned on its side,
yielding a hyperbolic-perturbed S-Schwarzschild.  It has the card
diagram structure of Fig.~\ref{SRN2fig}.

We can also perturb S-Schwarzschild by adding rods before
analytically continuing $z\to i\tau$, $M\to iM$.  We dub this the
`elliptic-perturbed S-Schwarzschild.'  It is different than
hyperbolic-perturbed S-Schwarzschild, and has the card diagram
structure of Fig.~\ref{chargedSbranecard}.  Turned on its side,
it yields an elliptic-perturbed Witten bubble, with the card
diagram structure of Fig.~\ref{bubblecard1}.  It is different
than the hyperbolic-perturbed Witten bubble.

In any of the cases, we can choose to analytically continue the mass
parameters of the additional rods or not.  Additionally, we can
displace some rods in the imaginary $z$-direction which affects
the $\tau$-center of their disturbance.  If we do everything in an
even fashion, i.e. we respect ${\rm Im}\,\tau\to-{\rm Im}\,\tau$,
the resulting geometry (at real $\tau$) will be real.  In
particular, rotating a rod at $z>0$ counterclockwise means
rotating its image at $z<0$ clockwise.  We see in the discussion
of the 2-rod example \cite{Jones:2004pz}
that there may be several choices for branches.

The card diagram techniques allow us to easily construct
these two inequivalent families of perturbed Witten bubble and
perturbed S-Schwarzschild solutions.  These and
other multi-rod, S-dihole, and infinite-periodic-universe solutions
are described in \cite{Jones:2004pz} and cannot be easily described
or understood without Weyl coordinates and the construction and
language of card diagrams.  The nontrivial $z\to i\tau$ continuation
is essential.

\section{Appendix: Electrostatic Weyl formalism} \label{EWeylapp}

The formalism of \cite{EmparanWK} can be extended for general $D$
to include an electrostatic potential.  This is somewhat
surprising since the electromagnetic energy-momentum tensor
$$T_{\mu\nu}=F_{\mu\rho}F_\nu{}^\rho-{1\over 4}g_{\mu\nu}F^2$$
is traceless only in $D=4$ and so Einstein's equations are more
complicated. Nevertheless, a cancellation does occur and one may
sum the diagonal Killing frame components of the Ricci tensor to
achieve a harmonic condition.

Follow the notation of \cite{EmparanWK} and add a 1-form potential
$A(Z,\Zbar)dt$ where $t=x^1$ is timelike ($\epsilon_1=-1$) and all
other $x^i$, $i=2,\ldots,D-2$ are spacelike (with
$\epsilon_i=+1$).  The metric takes the form
$$ds^2=-e^{2U_1}dt^2+\sum_{i=2}^{D-2}e^{2U_i}(dx^i)^2+e^{2C}dZd\Zbar,$$
from which we extract the frame metric
$$g_{\hat\mu\hat\nu}={\rm
diag}(-1,+1,\ldots,+1)\oplus\left[\begin{array}{cc}0&1/2\\
1/2&0\end{array}\right].$$ For $F=dA$ we have $F_{\hat Z\hat
t}=-F_{\hat t\hat Z}=\partial A\, e^{-U_1-C}$ and
$F_{\hat{\Zbar}\hat t}=-F_{\hat t\hat{\Zbar}}=\overline{\partial}
A\, e^{-U_1-C}$, all other components vanishing.  We compute
$F^2=-8\partial A\overline{\partial}A\,e^{-2U_1-2C}$ and
\begin{eqnarray*}
T_{\hat t\hat t}&=&2\partial A\overline{\partial}A\,e^{-2U_1-2C}\\
T_{\hat i\hat i}&=&2\partial
A\overline{\partial}A\,e^{-2U_1-2C}\quad(i\neq
1)\\
T_{\hat Z\hat Z}&=&-(\partial A)^2 e^{-2U_1-2C}\\
T_{\hat{\Zbar}\hat{\Zbar}}&=&\overline{T_{\hat Z\hat Z}}\\
T_{\hat Z\hat{\Zbar}}&=&0.
\end{eqnarray*}
The field equations are $R_{\hat\mu\hat\nu}-{1\over
2}g_{\hat\mu\hat\nu}R=T_{\hat\mu\hat\nu}$; taking the trace, we
get
$$R=-{4(D-4)\over D-2}\partial A\overline{\partial}A\,e^{-2U_1-2C}$$
and Einstein's equations are then
\begin{equation}\label{neweinstein}
R_{\hat\mu\hat\nu}=T_{\hat\mu\hat\nu}-{2(D-4)\over
D-2}g_{\hat\mu\hat\nu}
\partial A\overline{\partial}A\,e^{-2U_1-2C}.
\end{equation}
Form the sum $\sum_{i=1}^{D-2} R_{\hat i\hat i}\epsilon_i$; the
right side of (\ref{neweinstein}) gives
$$(D-4)2\partial A\overline{\partial}A\,e^{-2U_1-2C}-{2(D-4)\over D-2}(D-2)
\partial A\overline{\partial}A\,e^{-2U_1-2C}=0.$$
Hence (following (2.4)-(2.5) of \cite{EmparanWK}) we get
$$\partial\overline{\partial}\exp\left(\sum_{i=1}^{D-2}U_i\right)=0,$$
the Weyl harmonic condition.

One can add magnetostatic potentials along spatial Killing
directions as well.  We skip remaining details and give the
equations.  Let us assume $x^1$ is timelike and $x^i$ are
spacelike for $i=2,\ldots,D-2$, the potential is $A_{\it
1}=\sum_{i=1}^{D-2}A_i dx^i$, and the metric is
$ds^2=-e^{2U_1}(dx^1)^2+\sum_{i=2}^{D-2}e^{-2U_i}(dx^i)^2
+e^{2\nu}(d\rho^2+dz^2)$ and $w=\rho+iz$, $\partial_w={1\over
2}(\partial_\rho-i\partial_z)$. Einstein's equations are
\begin{eqnarray*}
\Delta U_1&=&{1\over 2}\Big(\sum_{i=1}^{D-2}(\nabla
A_i)^2e^{-2U_i}
+{D-4\over D-2}\sum_{i=1}^{D-2}(\nabla A_j)^2 e^{-2U_i},\\
\Delta U_k&=&{1\over 2}\Big(-(\nabla A_1)^2e^{-2U_1}-(\nabla
A_k)^2
e^{-2U_k}+\sum_{i\neq k,1}(\nabla A_i)^2 e^{-2U_i}\\
&&\qquad-{D-4\over D-2}\sum_{i\neq 1}(\nabla A_i)^2 e^{-2U_i}
+{D-4\over D-2}(\nabla A_1)^2 e^{-2U_1}\Big),
\end{eqnarray*}
and
\begin{eqnarray*}
\partial_w \sum_{i=1}^{D-2}\nu =-2\rho(\sum_{i<j}\partial_w U_i \partial_w
U_j +{(\partial_w A_1)^2 e^{-2U_1}\over
2}-\sum_{i=2}^{D-2}{(\partial_w A_i)^2\over 2}e^{-2U_i}).
\end{eqnarray*}
Maxwell's equations are
$$\nabla\cdot\big(\nabla A_i e^{-2U_i}\big).$$
All Laplacians and divergences are with respect to a flat 3d
axisymmetric auxiliary space with coordinates $\rho,z$.

\end{document}